\documentclass[preprint,12pt]{aastex}

\shorttitle{Envelope of L1527}
\shortauthors{Tobin et al.}

\newcommand{\msun}{\mbox{$M_{\sun}$}}
\newcommand{\rsun}{\mbox{$R_{\sun}$}}

\begin{document}

\title{Constraining the Envelope Structure of L1527 IRS: Infrared Scattered
Light Modeling}
\author{John J. Tobin\altaffilmark{1}, Lee Hartmann\altaffilmark{1},
Nuria Calvet\altaffilmark{1} \& Paola D'Alessio\altaffilmark{2}}
\altaffiltext{1}{Department of Astronomy, University of Michigan, Ann
Arbor, MI 48109; jjtobin@umich.edu}
\altaffiltext{2}{Centro de Radioastronom{\'\i}a y Astrof{\'\i}sica, UNAM, Apartado Postal
3-72 (Xangari), 58089 Morelia, Michoac\'an, M\'exico}

\begin{abstract}

We model \textit{Spitzer Space Telescope} observations of the Taurus
Class 0 protostar
L1527 IRS (IRAS 04368+2557) to provide constraints on its
protostellar envelope structure.  The nearly edge-on inclination of L1527 IRS,
coupled with the highly spatially-resolved near to mid-infrared images of this object
and the
detailed IRS spectrum, enable us to constrain the outflow cavity
geometry quite
well, reducing uncertainties in the other derived parameters.  The
mid-infrared scattered light image shows a bright central source within
a dark lane;
the aspect ratio of this dark lane is such that it appears highly unlikely
to be a disk shadow. In modeling this dark lane, we conclude that L1527 IRS is probably not
described by a standard TSC envelope with simple bipolar cavities. We find it necessary to model
 the dark lane and central source as a modified
inner envelope structure. This structure may be due either to a complex
wind-envelope interaction or induced by the central binary.
To fit the
overall SED, we require the central source to have a large near to mid-infrared
excess,
suggesting substantial disk accretion.  Our model reproduces the overall
morphology
and surface brightness distribution of L1527 IRS fairly well, given the
limitations of using axisymmetric models to fit the non-axisymmetric
real object,
and the derived envelope infall rates are in reasonable agreement with
some other investigations. IRAC observations of L1527 IRS taken 12 months
apart show variability in total flux and variability in the opposing
bipolar cavities, suggesting asymmetric variations in accretion. We also provide
model images at high resolution for comparison to future observations with
current ground-based instrumentation and future space-based telescopes.

\end{abstract}


\section{Introduction}

The \textit{Spitzer Space Telescope} has given us unprecedented views
of star forming regions. With \textit{Spitzer}, we are able to study
the earliest stages of star formation with far greater resolution,
sensitivity, and through higher extinctions than was previously
possible with ground-based near-infrared (NIR) imaging. These capabilities
of \textit{Spitzer} enable us to test theories of star formation
\citep[e.g.][]{tsc1984,shu1987} in detail using radiative transfer
models \citep[e.g.][]{whitney2003a}. The \citet[][TSC]{tsc1984} model,
describes the collapse of an initially spherical envelope which becomes
rotationally flattened as it collapses. This model has been the standard,
most widely used, model describing infalling protostellar envelopes for
the past two decades \citep[e.g.][]{adams1987, kch1993,osorio2003,whitney2003a, rob2006, furlan2007, tobin2007}.

Protostars drive powerful, bipolar molecular outflows which carve out
cavities in the circumstellar envelope. These cavities can be observed as
scattered light nebulae. The TSC envelopes alone predicted too little flux short ward of
10$\mu$m, necessitating the inclusion of bipolar outflow cavities into the
standard model \citep{calvet1994}. Thus far, a simple bipolar cavity structure carved out
of a TSC envelope has sufficed for modeling images and spectral energy
distributions (SEDs) of evolved, Class I, protostars. However, recent
observations by \textit{Spitzer} are enabling us to perform detailed
studies of outflow cavities and envelopes of the youngest protostars,
the Class 0 objects \citep{andre1993}. In particular, \textit{Spitzer} 
observations of outflow cavities in scattered light
permit more detailed analysis of the inner structure of protostellar
envelopes.

Observations of protostars with the Infrared Array Camera (IRAC)
\citep{fazio2004} have revealed numerous, prominent scattered light
nebulae associated with Class 0 objects \citep[e.g.][]{noriega2004,
tobin2007, seale2007}. Scattered light nebulae are also observed  in J,
H, and Ks (JHK) bands. However, the scattered light in JHK bands is
highly attenuated by the dense circumstellar envelope, therefore scattered
light nebula are often observed to appear brightest in the IRAC channels
numbered 1 to 4 (3.6, 4.5, 5.8, and 8.0$\mu$m) \citep{tobin2007}. 

Scattered light images are an important, alternative method of probing the envelope
structure of protostars, complementary to submillimeter and millimeter
studies of optically-thin dust emission. IRAC images are important to
scattered light studies because we can more clearly observe the envelope
structure since the dust is less optically thick than at JHK bands. Scattered light images 
enable clear observation of the outflow cavities and
may directly show the effects of the outflow on the envelope. In addition,
submillimeter images tend to be elongated along the outflow cavities due to heating
from the central protostar and disk. This additional emission complicates analysis of
the envelope structure at long wavelengths without knowledge of the outflow
cavity structure from scattered light images.

We cannot derive most physical parameters from scattered light images alone. Radiative transfer
modeling, in conjunction with a detailed observational analysis, must
be performed to derive the likely physical parameters of an object. The
results from radiative transfer modeling can yield insights into the
structure of protostars that scattered light images or dust continuum
observations alone would not reveal. Observations of L1527 IRS (IRAS
04368+2557) with the \textit{Spitzer Space Telescope} are an ideal starting point for
testing models of low-mass star formation.

The L1527 dark cloud is located within the Taurus molecular
cloud at a distance of $\sim$140pc. This object has been observed
extensively from the NIR to centimeter wavelengths
\citep{ohashi1997,chandler2000,loinard2002,hartmann2005}. L1527 IRS, hereafter L1527,
is classified as a borderline Class 0/I object with a large,
dense circumstellar envelope \citep{chen1995, motte2001}, we will refer 
to L1527 as a Class 0 object.  A unique property of L1527 is that
it is observed at a nearly edge-on ($\sim$90$^{\circ}$) inclination
\citep{ohashi1997}. The tightly constrained inclination makes L1527 ideal for modeling since
the inclination-dependent degeneracies on other modeled parameters 
(e.g luminosity, opening angle, infall rate) \citep{whitney2003b} are
limited. Also, since L1527 is relatively nearby, the spatial resolution
of the observations is about 2 times better than observations of protostars
 in Perseus and 3 times better than Orion.

The envelope properties of L1527 have been modeled
previously. \citet{kch1993} (KCH93) used TSC envelopes to model the SED
from IRAS; \citet{furlan2007} (F07) used a similar, improved model to
fit the SED and MIR spectrum of L1527. Finally, \citet{rob2007}
(R07) modeled L1527 using a large grid of TSC models calculated by
the \citet{whitney2003a} code \citep{rob2006}. Here we revisit L1527,
performing a more comprehensive observational analysis using scattered
light images, MIR spectra and photometry from the NIR
to millimeter wavelengths. We use these data combined to construct a new,
detailed model describing the envelope structure of L1527 in order to
test star formation theories and gain a more clear picture of protostellar
structure.

\section{Observations and data reduction}

\subsection{Spitzer Space Telescope Data}

Our data are taken from previous observations of Taurus in the
\textit{Spitzer} archive, processed with pipeline
version S14.0.0. L1527 has been observed twice; first on 07
March 2004 as part of the guaranteed time observer (GTO) survey of known
young stars in Taurus \citep{hartmann2005} and again on 23 February
2005 as part of the Taurus Legacy survey \citep{padgett2006}. The GTO
observations were performed in a 3 position, cycled dither pattern, with
a frame time of 12 seconds. The Legacy data mapped the entire Taurus
region twice with single frames of 12 seconds each and spatial limited
overlap. The legacy data of L1527 are comprised of two frames taken a
few hours apart, which require manual processing. We re-mosaicked 
the data using MOPEX version 061507
to make combined images from both observations and to create a more
clean mosaic of the Taurus Legacy data. 

The Taurus legacy survey also observed L1527 with the Multiband Imaging
Photometer (MIPS) \citep{rieke2004}. The observations were carried out
in mapping mode on 05 March 2005 in two separate observations 
of all three bands: 24, 70, and 160$\mu$m. We used the post-BCD data for
the 24$\mu$m observations; it was necessary to re-mosaic the 70$\mu$m
data using filtered BCD images because the post-BCD data have gaps due
to only half of the detector functioning. The 160 $\mu$m data did not
fully map L1527 and we could not use the data for photometry.

L1527 has also been observed by the Infrared Spectrometer (IRS)
\citep{irs2004} onboard the \textit{Spitzer Space Telescope}. The
spectrum was taken in both orders of the short-low (5 - 14$\mu$m) and long-low (14 - 40$\mu$m) modules
on 27 February 2004. In our analysis, we use the same spectrum
as presented in F07. The spectral resolution is R $\sim$
120-60 for both modules.

\subsection{Near-Infrared Observations}

We observed L1527 at the MDM Observatory at Kitt Peak on 27 and 29 December 2007
using the 2.4m Hiltner telescope with the NIR imager TIFKAM \citep{tifkam}. We observed
L1527 in the J, H, Ks, and H$_2$ (2.12$\mu$m) filters on 27 December and the H$_2$ continuum (2.09$\mu$m) filter
on 29 December; with total integration times in J and H-band of 10 minutes and
a total integration time of 35 minutes in Ks, H$_2$, and H$_2$ continuum bands. The H$_2$ continuum filter
is a line-free, narrowband filter with a bandwidth matching the H$_2$ filter. 

The observations were conducted in a
5 point dither pattern with 1 minute of integration per dither position. After each complete 
dither pattern, we chopped off source and observed the sky in the same 5 point dither pattern to
construct a median sky. The off-source imaging was necessary because TIFKAM has a 3$^{\prime}$ field of view
and L1527 nearly fills the field. The data were reduced using standard routines from the 
upsqiid package for flat fielding, sky subtraction, and combining. Our J-band image does not
detect L1527 and the H-band image only marginally detects L1527. However, L1527 is clearly observed in
 the Ks-band image. The H$_2$ image marginally detects L1527 with comparable signal-to-noise (S/N)
to H-band, while the H$_2$ continuum band image does not detect L1527.

\subsection{Additional Infrared Data}

We also included mid-infrared data from the Infrared Space Observatory
(ISO) in our study. We used ISOCAM data from the 6.7, 9.6, 11.3 and 14.3$\mu$m bands observed
on 02 October 1997. These bands are overlapped by the IRS instrument,
however, we used these images to constrain any extended emission at
these wavelengths. We used pipeline processed data from the ISO science
archive, performing no further reduction of these data.

Finally, we used data from the Two Micron All Sky Survey (2MASS)
\citep{2mass} in our study of L1527. We use the calibrated data frames
for the L1527 region from the 2MASS archive.

\subsection{Photometry}

We performed aperture photometry on L1527 using the Image Reduction and
Analysis Facility (IRAF)\footnote{IRAF is distributed by the National
Optical Astronomy Observatories,
which are operated by the Association of Universities for Research
in Astronomy, Inc., under cooperative agreement with the National
Science Foundation.}. Photometry was not straightforward
since a background annulus cannot be used because it includes extended
emission surrounding the source. Instead, using the `imstat' procedure
in IRAF, we measured a large area of sky adjacent the source that was
devoid of emission to obtain an average background value per pixel. This
value was then multiplied by the aperture area measured by IRAF and
subtracted from the photometric value of the source. Also, local
extinction of the background will result in over-subtraction of the
background on the source. This over subtraction is greatest in IRAC
channels 3 and 4 where the background interstellar medium (ISM) and
zodiacal light emission are highest.

We measured the flux within 1000 and 10000 AU aperture radii centered
on the position of L1527; $\alpha=$ 04:39:53.9 $\delta=$ +26:03.09.6
(J2000).  These apertures correspond to 6 and 60 pixel radii at a distance
of 140 pc and the IRAC pixel scale of $1\farcs2/\rm{pixel}$.
 We use two apertures in our analysis because
the 1000 AU aperture probes emission on small scales, dominated by the
central envelope structure and the 10000 AU aperture probes emission
dominated by extended scattered light in the outflow cavity. Lastly, we applied
aperture corrections to our photometry using the methods of extended
source calibration described on the \textit{Spitzer Science Center}
website. Photometry and aperture corrections are listed in Table 1.

For the ISO data, we used the same procedure as above, however
we were only able to use apertures of 1000 (7$\farcs$2) and 5000 AU
(36$^{\prime\prime}$) due to the limited field of view. The ISO photometry
may have up to 30$\%$ errors as aperture corrections were unknown and
background subtractions are uncertain.

The 2MASS photometry was performed using the same procedure as the IRAC data. The
pixel size of the 2MASS detectors was 2$\farcs$0/pixel so our apertures
were 3.57 pixels for 1000 AU and 35.7 pixels for 10000 AU. 

Finally, photometry from the TIFKAM data were also taken using the method for IRAC data.
The pixel size on the TIFKAM detector corresponds to 0$\farcs$2/pixel, translating
to a 35.7 pixel aperture for 1000 AU and a 357 pixel aperture for 10000AU. The data
were calibrated using the faint NIR standards listed in \citep{hunt1998}.

\section{Observed Morphology}

\begin{figure}

\plotone{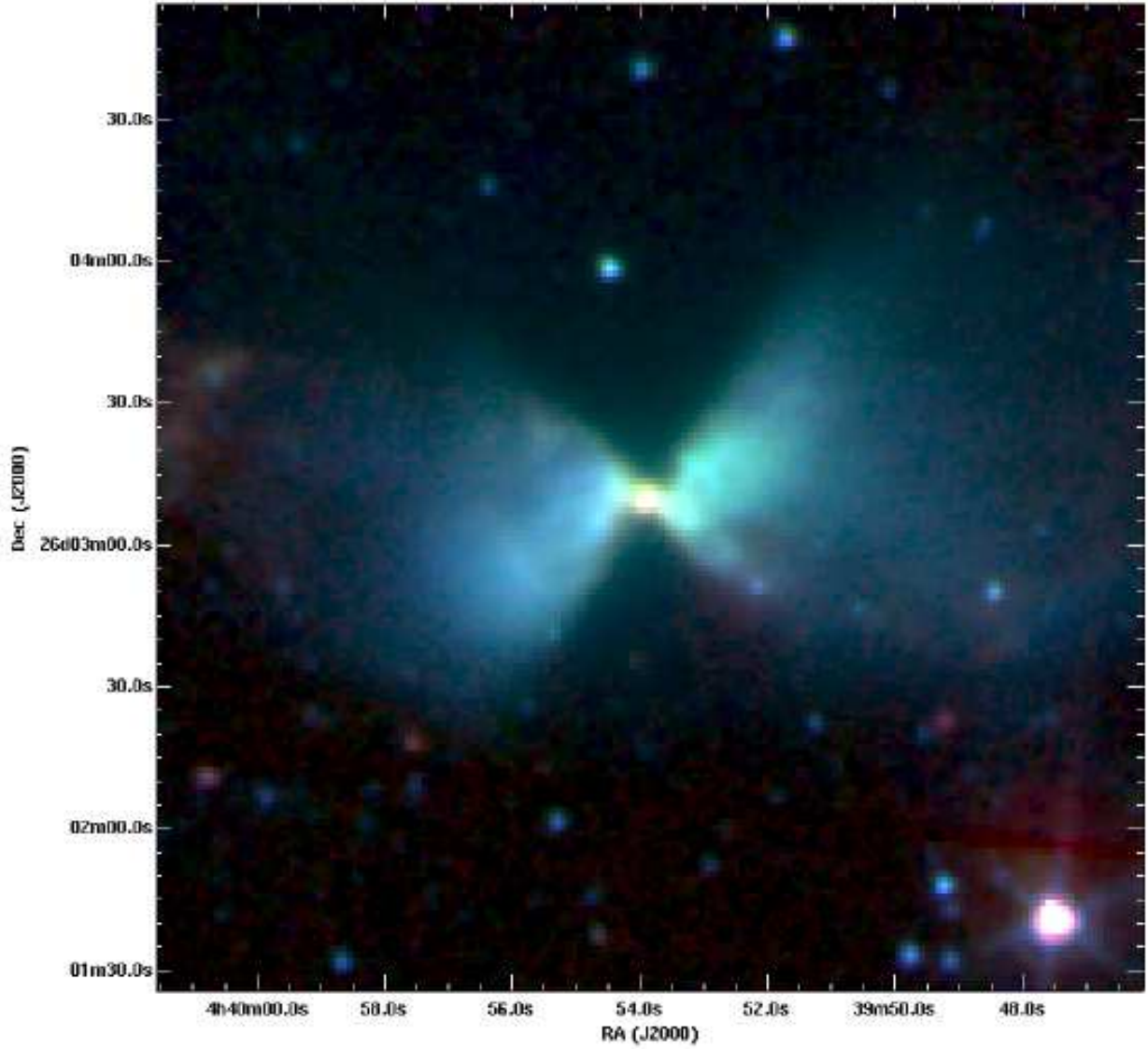}
\caption{False-color IRAC image of L1527 IRS (IRAS 04368+2557), Blue: 3.6$\mu$m,
Green: 4.5$\mu$m, and Red: 8.0$\mu$m. Image box is 200$^{\prime\prime}$
on a side, corresponding to $\sim$28000 AU at a distance of 140pc.}
\label{color}
\end{figure}

\begin{figure}
\begin{center}
\includegraphics[angle=-90, scale=0.75]{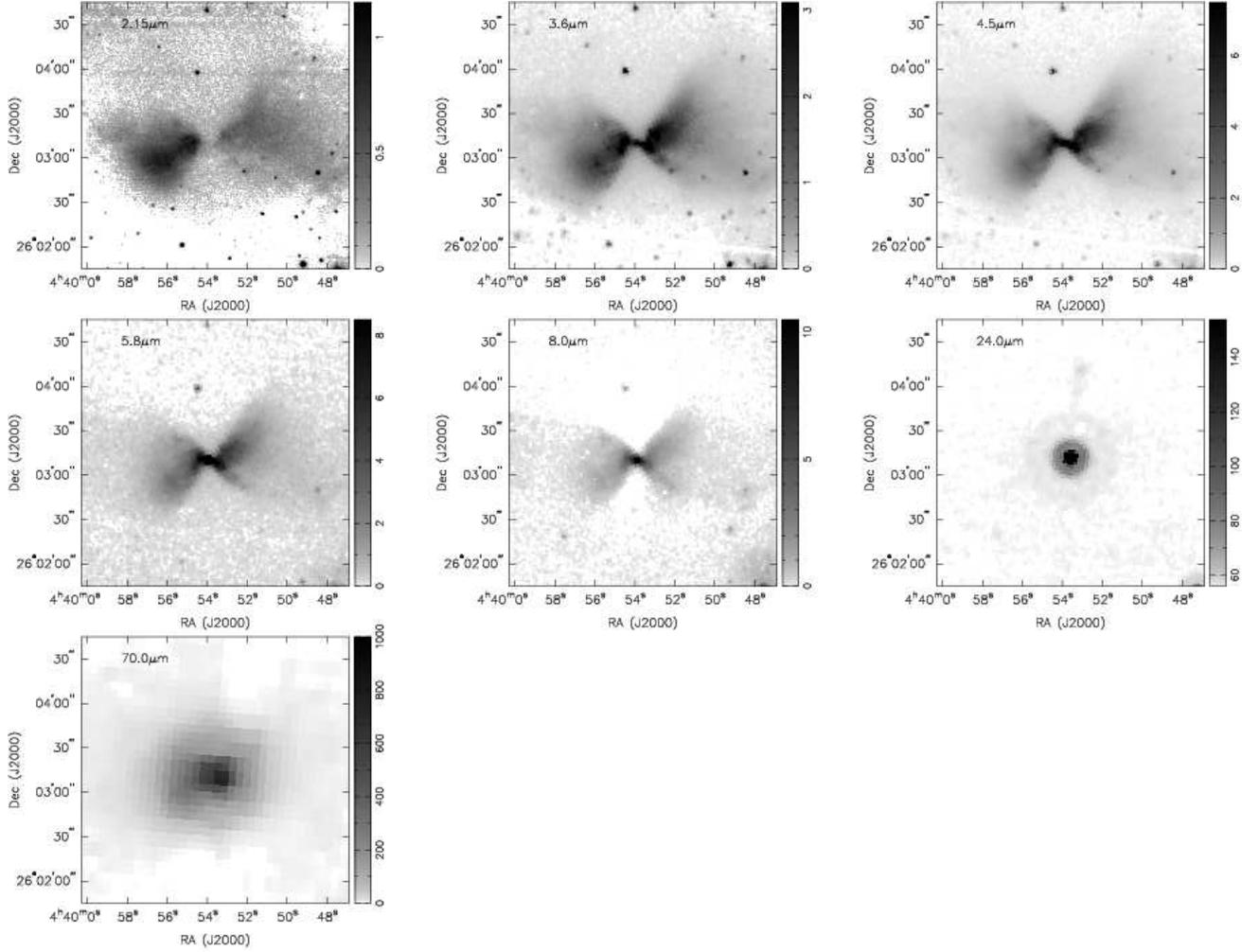}
\end{center}
\caption{Near-infrared and \textit{Spitzer} images of L1527 as observed by TIFKAM, IRAC, 
and MIPS. Images are 180$^{\prime\prime}$ on a side, corresponding to $\sim$25000 AU at
a distance of 140pc. Units of intensity are MJy/sr.}
\label{4chan}
\end{figure}

As shown in Figs. \ref{color} and \ref{4chan}, L1527 exhibits bright
bipolar scattered light nebulae which extend roughly
1.5 arcminutes (12600 AU) on each side of the central source. 
In addition to the large, bright cavities, the most striking features are the `neck' and absorption lane
between the reflection nebulae. The absorption lane is clearly shown in the
Ks-band image (Fig. \ref{4chan}), having a width of $\sim$10$^{\prime\prime}$ (1500 AU). 
The `neck' becomes apparent in the IRAC images as
a thin strip of emission bridging the gap between the cavities. The `neck'
is shown in more detail in Fig. \ref{zoom}.

In addition, we observe a bright source between the cavities on the `neck'
in the IRAC images, see Fig. \ref{zoom}. This feature was not detected in the
TIFKAM or 2MASS observations. This source appears to be slightly extended as compared to
a point source in all IRAC channels, though the surrounding nebulosity
makes it difficult to be certain. We will discuss the possible nature
of this object in \S5.

In the MIPS 24$\mu$m images, L1527 is a point source, but at 70$\mu$m there
is marginally resolved structure; L1527 appears elongated along the outflow axis, see Fig. \ref{4chan}.
The radius of the envelope at 70$\mu$m is about 70$^{\prime\prime}$ or about 10000AU.
 L1527 was not completely 
mapped at 160$\mu$m, but we again observe extension along the outflow axis. Also,
we estimate that the L1527 envelope has a radius of approximately 100$^{\prime\prime}$
or 14000 AU from the 160$\mu$m image.

While the bipolar nebulae of L1527 may in general seem prototypical and
simple to model, in reality L1527 is very complex. We noticed that in
IRAC Channels 1 and 2 there is an inherent lumpiness to the scattered
light cavities presumably from substructure in the cavity walls; the
substructure becomes unresolved in IRAC channels 3 and 4. In addition
to the lumpiness of the scattered light, L1527 shows several asymmetries
in brightness and geometry.

\subsection{Asymmetry}

The position angle (PA) of the outflow from L1527 is $\sim$90$^{\circ}$ east of
north \citep{zhou1996, hoger1998}. Also, we observe faint Herbig-Haro objects
at $\sim$230$^{\prime\prime}$ west and $\sim$160$^{\prime\prime}$
east of L1527 in the IRAC images (Fig. \ref{HH}); they are consistent with a PA of
$\sim$90$^{\circ}$. 
Using this PA as the symmetry axis for the outflow
cavities, there is clear asymmetry which is quite plain in Figs. \ref{color} and \ref{4chan}. The
southeast side of the cavity opens wider than the northeast and vice
versa. Also, the cavity has an azimuthal brightness asymmetry. In the east
side cavity, the south side is brighter that the north side and in the
west side cavity the north side is brighter than the south side. The asymmetry seems to be a
global effect on the entire envelope since the departures from spherical
symmetry are observed to be qualitatively point symmetric. Also, the asymmetry appears to be 
present at many different wavelengths, even in the submillimeter images, see Fig. \ref{modelscuba}. 
The submillimeter contours are overall
boxy in shape. In the southeast and northwest of Fig. \ref{modelscuba}, the contours are more
extended than the northeast and southeast contours, looking like a parallelogram.

\subsection{Variability}

We observe the cavities of L1527 to have substantial brightness
variations over time. Comparing the two epochs of observations in
Fig. \ref{2epochs}, we can see that both sides of the cavity
exhibit independent variability over the entire cavity. 
In 2004 the western cavity was
much brighter than in 2005; and in 2004 the eastern cavity is fainter
than in 2005.

Fig. \ref{sectors} shows the flux averaged over sectors of annuli
spaced radially for each epoch. The variation
between the two epochs
is not large enough to affect modeling and the brightness of the cavities is fairly flat close
to the center. The flatness is the result of the curved cavity shape;
the distance of the cavity wall to the illuminating source changes
slowly in this region.  Also, at a radial distance of greater than
10$^{\prime\prime}$ the cavity brightness falls off roughly $\propto$ R$^{-2}$
in both epochs, consistent with scattered light from a central
point source incident on an optically thick wall \citep{whitney1993}. The
bump at about 30$^{\prime\prime}$ is due to a brightened cavity feature
present in both cavities at this distance. IRAC Channels 1 and 2 are very similar in their ratios over the
entire measured region, see Fig. \ref{2epochs}. IRAC Channels 3 and 4 are similar out to about
50 arcseconds; where the scattered light falls off in these channels. IRAC channels
3 and 4 are not shown in Fig. \ref{2epochs} as the divided images are very noisy.

\subsection{Local Extinction}
\subsubsection{8$\mu$m Absorption}
With the data from \textit{Spitzer}, we are not only observing the envelope
structure through scattered light, we also directly probe the envelope
structure in absorption at 8$\mu$m. IRAC Channel 4 is the widest IRAC
band with a bandwidth of 2.93$\mu$m, extending from 6.44 to 9.38$\mu$m,
overlapping with the 6.85$\mu$m organic ice, 9.0$\mu$m ammonia ice, and
most importantly the 9.7$\mu$m silicate feature. The presence of these
absorption features in IRAC Channel 4 and a significant interstellar
background make it an ideal band to observe MIR extinction. Observing
absorption in the MIR is not new, this type of analysis has been
performed extensively on infrared dark clouds located in the galactic
midplane \citep[e.g.][]{simon2006, ragan2006}. Also, \citet{looney2008}
observed L1157 to have a flattened envelope in 8$\mu$m absorption.

To turn our IRAC 8.0$\mu$m image into an optical depth image, we first
subtracted the estimated zodiacal light; a model calculated value is found
in the image header. Then we measure
the residual background emission, using the same method we used for the
photometry background measurement. The 8.0$\mu$m image is then divided by
the background measurement and the natural log of each pixel is calculated,
yielding an optical depth image. This method assumes a constant background 
which may not be a realistic assumption. However, given the low S/N of the 
background and small angular size of L1527, modeling of the background as described in \citet{simon2006}, would
not improve the analysis.

Fig. \ref{HH} shows the 8$\mu$m optical depth contours overlaid on the
IRAC 8.0$\mu$m image. In addition to the asymmetries observed in scattered
light, the 8.0$\mu$m absorption is asymmetric. The north side of the
envelope has significantly more optical depth and covers a larger area than the south.
This observation infers that there is more material to the north in
our line of sight, in agreement with submillimeter observations
shown in Fig. \ref{modelscuba}. The submillimeter contours are more
extended on the north side of L1527 by about 20 to 30$^{\prime\prime}$
than on the south side.

Using these optical depth data, we can get a rough idea of the amount
of material in the envelope. Assuming the 8.0$\mu$m opacity from
\citet{draine2001} we derive a mass of 0.22 \msun. The measured mass is low compared 
to previous estimates of 2.4 $M_{\sun}$ to 0.8 $M_{\sun}$ from ammonia and dust emission
\citep{benson1989, shirley2000,motte2001}. However, there are large uncertainties
in the opacity (emissivity) in the MIR and at long wavelengths. Also, the level of 
residual zodiacal light contamination of the IRAC images is uncertain and in general 
our measurement is uncertain due to low S/N of the background. The measurement is also
at best a lower limit as scattered light emission is present over roughly half 
the envelope masking some of the extinction.

\subsubsection{Inclination Dependent Extinction}

Radiative transfer models \citep{whitney2003b} and simple geometry show
that the local extinction of the outflow cavity lobes will depend on line
of sight inclination. Local extinction refers to self-extinction by the
envelope, not the ambient molecular cloud. With constant illumination of
a spherically symmetric envelope, the blue-shifted outflow cavity (tilted
toward the observer) should appear brighter than the red-shifted cavity
(tilted away). The blue and red-shifted cavities are determined from
observations of rotational transitions of molecules at submillimeter
and millimeter wavelengths, e.g. CO, SiO, HCN, etc. For L1527, the
eastern cavity is blue-shifted and the western cavity is red-shifted
\citep{zhou1996, bontemps1996, ohashi1997, hoger1998}. There is some overlap in
the velocity channels indicating that L1527 is observed close to edge-on.

In the Ks-band image from TIFKAM, both the blue and red-shifted
cavities are visible. However, the blue-shifted cavity is brighter by about a factor of 2
in some regions. 2MASS observations only show the blue-shifted cavity.
In the two epochs of IRAC observation, the cavities have appeared roughly equal
in brightness in the 2005 epoch, and the blue-shifted cavity being brighter in the 2004 epoch.
These observations are consistent with dependence of cavity brightness on inclination
highlighted in \citep{whitney2003b}, indicating that L1527 is close to edge-on
but slightly inclined.

This investigation of the observed morphology yields many observational
constraints which were necessary to take into account while constructing
our model. Some constraints such as observing L1527 nearly edge-on
greatly helps modeling by reducing several degrees of freedom for
opening angle and inclination. The variability does not greatly hinder
the modeling effort since the overall morphology of the object remains
constant and the overall increase or decrease in intensity is at most
a factor of 2. Finally, reproducing the observed central source
and dark lane requires specific attention in constructing a model of L1527.

\section{Modeling}

\subsection{Envelope and Disk Structure}

To interpret the observations, we used the radiative transfer code
of \citet{whitney2003a} to construct a model of the L1527 data. This
code uses the Monte Carlo method to calculate radiative equilibrium
and include multiple scattering. The model has many parameters to tune,
some more important than others; we will review the parameters important
to our study.

We model the central protostar with a stellar atmosphere 
for a given effective temperature and radius.
In addition to the protostellar luminosity, we found
that we needed strong disk emission. The strong disk emission can come 
from high accretion luminosity, a bright star illuminating the disk, or
a combination of both.

We assume the typical model for T Tauri accretion in which material from the inner disk falls along
magnetic field lines onto the protostar creating an
an accretion shock which is assumed to radiate as a blackbody
over an area of 0.01 of the stellar surface \citep{calvet1998}. 
In addition to heating the envelope, emission from this accretion shock 
heats the disk along with the radiation from the stellar photosphere.
Local viscous dissipation also heats the disk, but 
the dominant disk heating is the protostar plus accretion shock irradiation.
These disk heating mechanisms combined result in the disk emitting substantial
NIR to MIR radiation.

The total system luminosity is
\begin{equation}
L_{total} = L_* + L_{shock} + L_{disk},
\end{equation}
where L$_{shock}$ is the luminosity of the accretion shock on the star,
\begin{equation}
L_{shock} = \frac{GM_*\dot{M}_{disk}}{R_{*}}
\left(1-\frac{R_*}{R_m}\right)
\end{equation}
and $L_{disk}$ is the luminosity from viscous energy release,
\begin{equation}
L_{disk} = \frac{GM_*\dot{M}_{disk}}{R_{in}}.
\end{equation}
In these equations, $M_*$ is the central source mass, $\dot{M}_{disk}$
is the disk accretion rate of the disk onto the star, $R_{in}$ is the inner
radius of the disk, generally the dust destruction radius where the dusty
disk is truncated, and $R_m$ is the magnetospheric truncation radius. In
reality, the disk extends to $R_m$; the disk inside the dust destruction
radius $R_{dd}$ is composed of dust-free gas which may be optically
thin or thick. Currently, there are few observational constraints on the
properties of the inner disk; thus, it is not included in as an energy
source in the model. The gaseous inner disk may be an important detail
that our model lacks.

The disk structure is assumed to be a standard flared accretion disk
structure \citep[e.g.][]{hartmann1998} with a density profile given by
\begin{equation}
\rho(r)=\rho_0 \left( 1- \left( \frac{R_*}{r}
\right)^{1/2} \right) \left( \frac{R_*}{r} \right)^{\alpha}
exp\left[-\frac{1}{2}\left(\frac{z}{h(r)}\right)^2\right]
\end{equation}
where $r$ is the radial coordinate of the disk, and $h(r) =
h_0(r/R_*)^{\beta}$. The model does not assume any temperature structure
for the disk \textit{a priori} thus the $h(r)$ is not calculated
self-consistently throughout the disk. The model solves the
hydrostatic equilibrium equation to obtain the scale height $h(r)$ at the
dust destruction radius, modified by the flaring
power-law $\beta$, for the entire disk structure. However, to ensure we
model parameters that are roughly physically consistent, we manually specify a scale height for
the model determined from the scale height at the dust destruction radius

\begin{equation} h_{dd} = \left( \frac{ k T_{dd} R_{dd}^{3} }{ G M_{*}
\mu  m_{p} } \right)^{1/2}\end{equation}
where $T_{dd}$ is the dust destruction temperature, 1400K, $\mu$ is the
atomic mass unit assumed to be 2, $m_p$ is the mass of a proton, $k$
is the Boltzmann constant, and $R_{dd}$ is the dust destruction radius
given by
\begin{equation}
R_{dd} = \left(\frac{T_{eff}}{T_{dd}}\right)^{2.1}.
\end{equation}
$T_{eff}$ refers to the effective temperature of the star, including
the accretion luminosity. The code states that the power 2.1 comes from
an empirical fit, if the wall and star were in equilibrium and assuming
perfect blackbody radiation, the power should be 2.

For the circumstellar envelope, we assume a TSC envelope with the density
structure given by
\begin{equation}
\rho(r) = \frac{ \dot{M}_{env} }{ 4\pi(GM_*r^3)^{1/2} }\left(1 + \frac{
\mu }{ \mu_{\circ} }\right) \left( \frac{ \mu }{ \mu_{\circ} } + \frac{
2\mu_{\circ}^2 R_c }{ r }\right)^{-1}.
\end{equation}
$\dot{M}_{env}$ is the mass infall rate of the envelope onto the disk,
$R_c$ is the centrifugal radius where rotation becomes important, $\mu$ =
cos $\theta$, and $\mu_0$ is the cosine polar angle of a streamline out to
r $\rightarrow$ $\infty$. Inside of $R_c$; $\rho_{env} \propto r^{-1/2}$
and outside $R_c$; $\rho_{env} \propto r^{-3/2}$. The centrifugal radius
controls the optical depth to the center of the envelope, increasing $R_c$
decreases the overall envelope density. The overall mass of the envelope
does not strongly depend on $R_c$ as the centrifugal radius is generally 
small compared to the outer envelope radius.

Bipolar cavities in the envelope
are obviously required to fit the observations of extended scattered
light and thermal emission.
These bipolar cavities extend from
the central protostar to the outer radius of the envelope. Curved or
streamline cavities were options in modeling, we adopted curved cavities
as they allowed for greater flexibility in modifying the cavity shape. The
curved cavity structure is defined by $z = C(x^2 + y^2)^{b/2}$, where $C$
is a constant determined by a relation between the cavity opening angle
and envelope radius, and $b$ is the shape parameter, the power of the
polynomial defining the cavity shape. In our model, 
it was necessary to use two cavities of different sizes in the inner
and outer envelopes. The cavity structure will be discussed further in 
\S5.

Table 2 lists the parameters of the model we found to best fit L1527.  The stellar parameters
are uncertain and were simply taken to be those of typical Taurus
T Tauri stars and were not varied. The total system luminosity (star plus disk accretion)
is very important, depending on both the fixed stellar and the accretion
disk parameters.  There are several parameters
for the accretion disk, but in terms of the modeling, the most important
ones are the disk accretion rate, which sets the disk and accretion shock luminosity, and
scale height specifying the inner disk ``wall'' at the dust destruction
radius, which dominates the emission in the IRAC bands (see \S5).
The other disk parameters are relatively unimportant.  The envelope
infall rate in conjunction with the assumed stellar mass set 
overall optical depth of the envelope and the amount of long-wavelength thermal emission.
The cavity properties are somewhat arbitrary but are simply tuned to
match the observed morphology.  The inclination angle is essentially edge-on,
and the cavity and ambient densities, sometimes used in other modeling, are
set to zero in our investigation.

\subsection{Dust Properties}

The default envelope dust properties assumed in the model are R$_V = 4$ dust
grains with 5\% H$_2$O ice coatings. This is one of the standard dust
opacity tables included with the \citet{whitney2003a} model which is
 derived from \citet{kmh1994}. These dust properties are
intended to be similar to the ISM dust in
Taurus. These dust grains do not include the 6.0$\mu$m H$_2$O,
6.85$\mu$m CH$_3$OH, and 15.2$\mu$m CO$_2$ ice features; we observe
all these ice features typically in Class 0 and Class I protostars
 (F07).

The dust properties are a source of uncertainty in the model. Dust
grain models appropriate for the ISM may not be appropriate
for Class 0 envelopes. If dust grains in the diffuse ISM of Taurus are
able to develop ice coatings \citep{whittet2001} and grow in dense
regions \citep{flah2007, zuninga2007}, this leads
to the possibility that considerable grain growth could occur in dense
protostellar envelopes. Larger grains would have higher albedo out
to longer wavelengths and could help to explain the intensity of the
observed scattered light in the MIR.

Driven by our observations, we have constructed a dust model which uses
larger dust grains than a standard ISM dust model. The dust model is calculated
using the method described in \citet{dalessio2001}, with additional 
optical constants for graphite from \citet{draine1984}. The grain size
distribution is defined by a power law grain size distribution $n(a) \propto a^{-3.5}$, with a$_{min}$
= 0.005$\mu$m and a$_{max}$ = 1.0$\mu$m. We used dust grains composed
of graphite $\zeta_{graph} = 0.0025$, silicates $\zeta_{sil} = 0.004$,
and water ice $\zeta_{ice} = 0.0005$; abundances are mass fractions relative to gas. The
given abundances infer a gas to dust ratio of 133. Our dust model does
not include organic molecules, all carbon is in the form of graphite, nor
 does it include the additional ice features detailed at the
start of this section. Dust grains
larger than a$_{max}$ = 1.0$\mu$m would have too much albedo long-ward of
8.0$\mu$m yielding significant extended emission between 8.0$\mu$m and
15$\mu$m. Observations by ISO do not show significant extended emission
above the background at 9.6 or 11.3$\mu$m. These observations constrain
the maximum possible size of dust grains in the outer envelope of L1527 to
be $\sim$1$\mu$m in radius. The opacity and albedo curves of our assumed dust model
is plotted against an ISM dust model \citep{kmh1994} in Fig. \ref{dust}.

\subsection{Model Fitting}

Due to the inherent uncertainties in modeling L1527 and observed
asymmetries, we did not quantitatively fit our images or SEDs using
a $\chi^2$ or equivalent minimization routine. To fit the observed
morphology of the image, we convolved the model image with the appropriate,  
instrument specific, PSF
and compared it to the observed data. If an image looked similar we noted the
particular parameter set and ran the model again with more photons to
obtain a higher S/N image.
We cannot duplicate the asymmetric
cavities of L1527, due to the axisymmetry of the model, but the general
morphology modeled is quite similar. 

With the high S/N images, we compared the intensity averaged over sectors
radially spaced (Fig. \ref{modelsectors}) to compare the data and the models. 
In Fig. \ref{modelsectors} we see that the 3.6$\mu$m channel agrees the best
while the overall strength of the scattered light in the 4.5,
5.8 and 8.0$\mu$m channels falls short at all radii, though the shape
of the curves is quite similar. At Ks-band, the model also roughly agrees with the observations.
Similar to the
IRAC data, the Ks-band plot also has the bump at 30$^{\prime\prime}$, probably due to a bright
spot in the cavity or an outflow knot. In addition, the intensity of the modeled central
object in the IRAC bands does roughly agree with the observations.

The SED fitting was done
by eye, adjusting the luminosity controlling parameters until the model was
as close to the observed SED as possible. We were able to nearly bring
the overall SED of the model and observations into agreement.  Also,
the asymmetry of the cavities does account for some of the discrepancy
between the model and observed SED in the 10000 AU aperture. Also, we apply a foreground extinction of
A$_V$ = 3 to the SED and images for fitting. The addition of extinction extinguishes the
extended scattered light the model predicts in the visible while hardly affecting 
the portion of the SED we are fitting. The extinction
toward L1527 is uncertain since it is a protostar embedded within a dark cloud. However,
the assumed extinction is not atypical for Taurus. 

\section{Results}

In order to match the observed morphology of L1527, changes to
the standard \citet{whitney2003a} model were necessary. The largest necessary change was
to the bipolar cavity configuration. 
Here we describe the changes to the standard
parameter set of the model and review the resulting physical parameters and their
significance.

\subsection{Reproducing the Scattered Light Morphology}

At the start of our modeling effort we adopted the parameters fit by
F07; see Table 2. We soon found that we could not replicate the dark lane,
`neck', or central object with the standard bipolar cavity geometry of the model and 
a relatively flat disk. In Fig. \ref{modelsingle} we show how the model looks using 
the default, single cavity structure. These images do show a dark lane, however it is 
very thin, and there is no central source in the IRAC images. The appearance of this dark lane is similar to 
those observed by \citet{padgett1999}, which are the
relatively thin shadows of dusty circumstellar
disks. The thickness of the dark lane in L1527 requires an structure that is as
wide as it is thick, unlike a disk. We thus concluded that the observed dark lane and `neck' 
must be due to the envelope structure and the standard bipolar cavities of the model would not suffice.

\begin{figure}
\begin{center}
\includegraphics[angle=-90, scale=0.7]{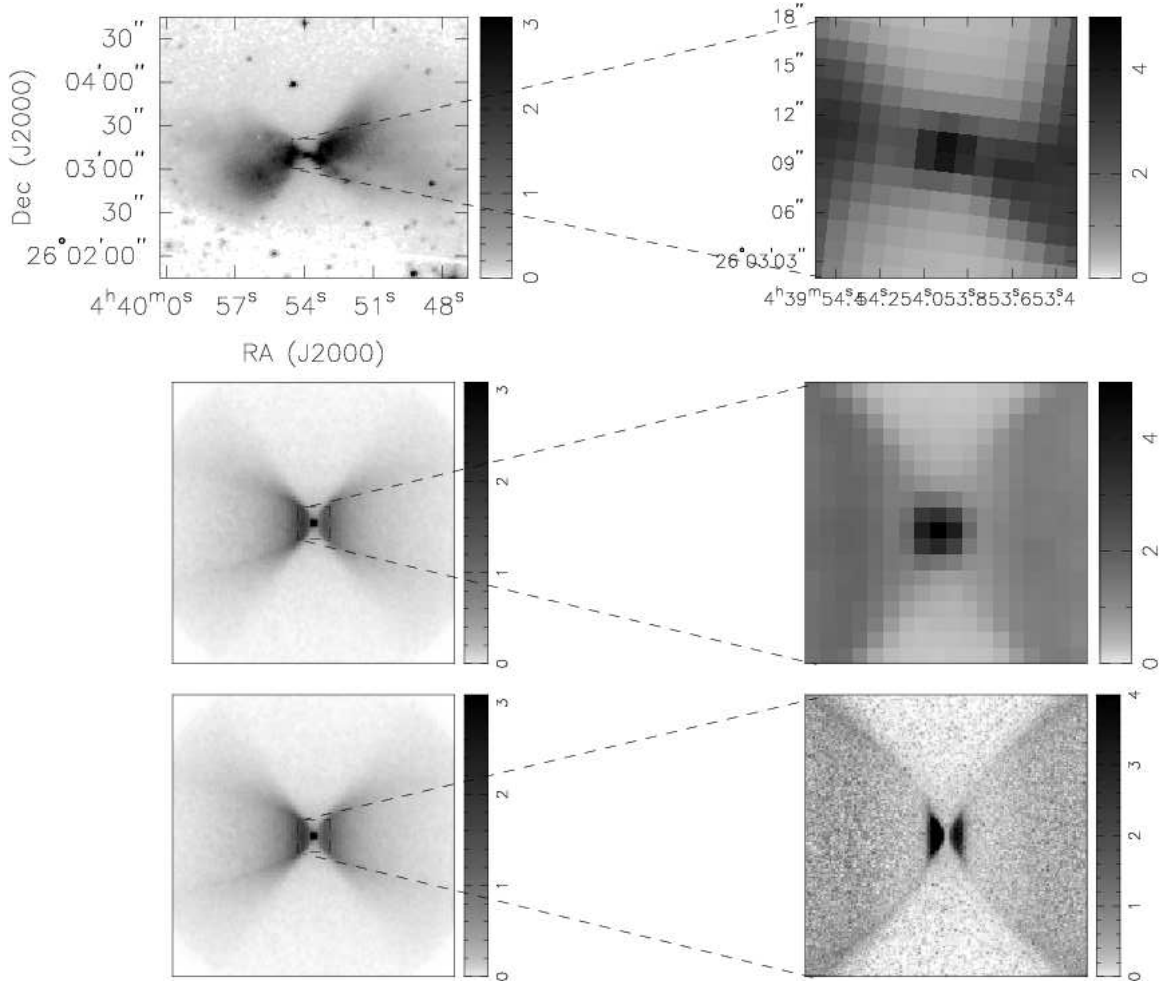}

\end{center}

\epsscale{0.8}
\caption{Top: IRAC 3.6$\mu$m image of L1527 zooming on on the `neck'
as described in \S3. Middle: Best fitting model image zooming in on
inner envelope structure. Bottom: Best fitting model image with 
unconvolved view of of the inner envelope. Units are MJy/sr. }
\label{zoom}
\end{figure}

\begin{figure}
\begin{center}
\includegraphics[angle=-90, scale=0.85]{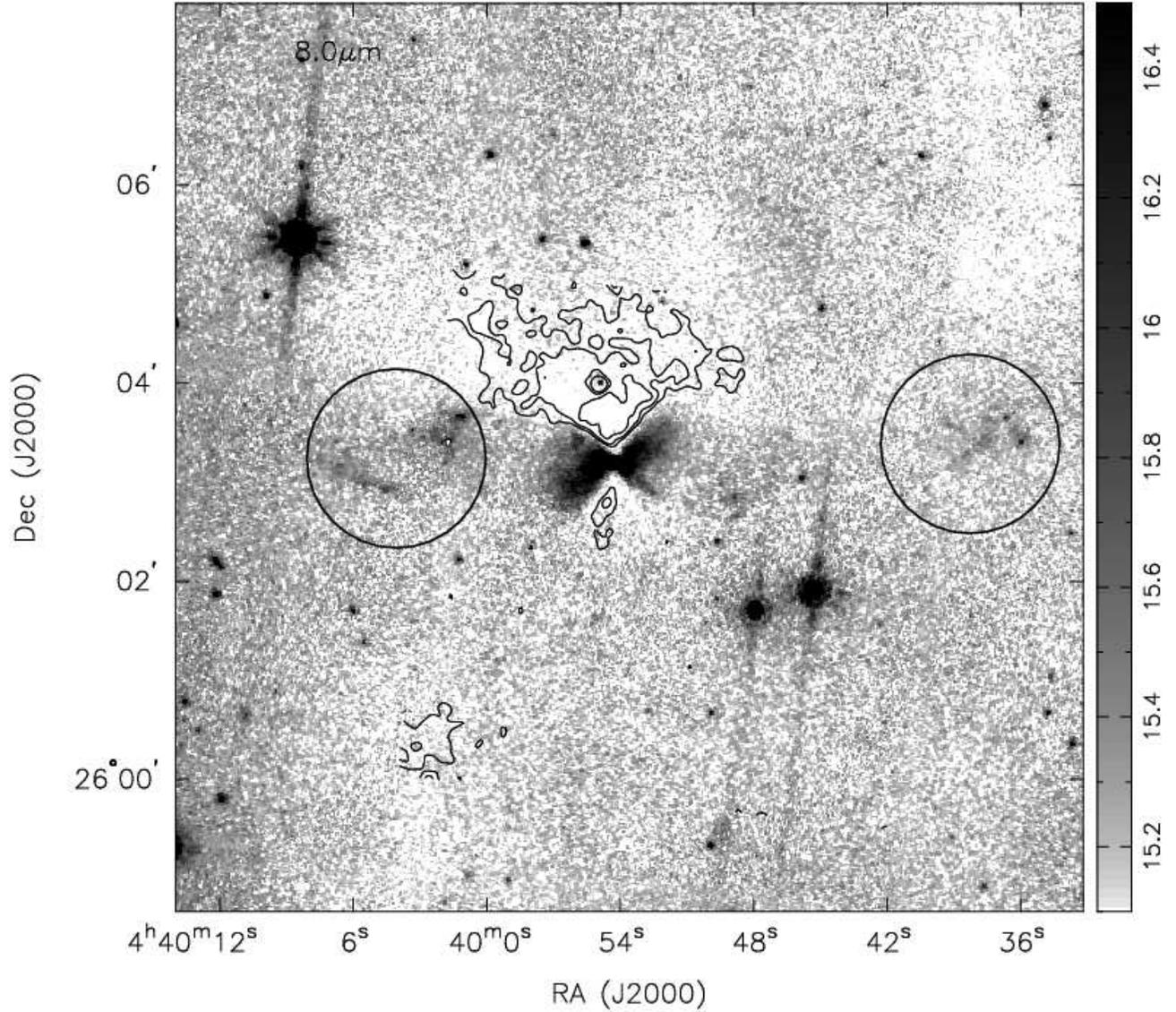}
\end{center}
\caption{Wide-field view of L1527 region at 8.0$\mu$m. Herbig-Haro objects in apparent 
association with L1527 are circled. These objects infer an outflow PA $\sim$90$^{\circ}$.  
Assuming an outflow velocity of 100km/s, these outflow knots have a dynamical age of only 1100 - 1600 yrs.
Contours are 8.0$\mu$m optical depth contours corresponding to $\tau$ = 0.08, 0.10, 0.126 and 0.16. The
contours match up with the regions where there is a lack of background emission. Units are MJy/sr. }
\label{HH}
\end{figure}

\begin{figure}
\begin{center}
\includegraphics[angle=-90,scale=0.75]{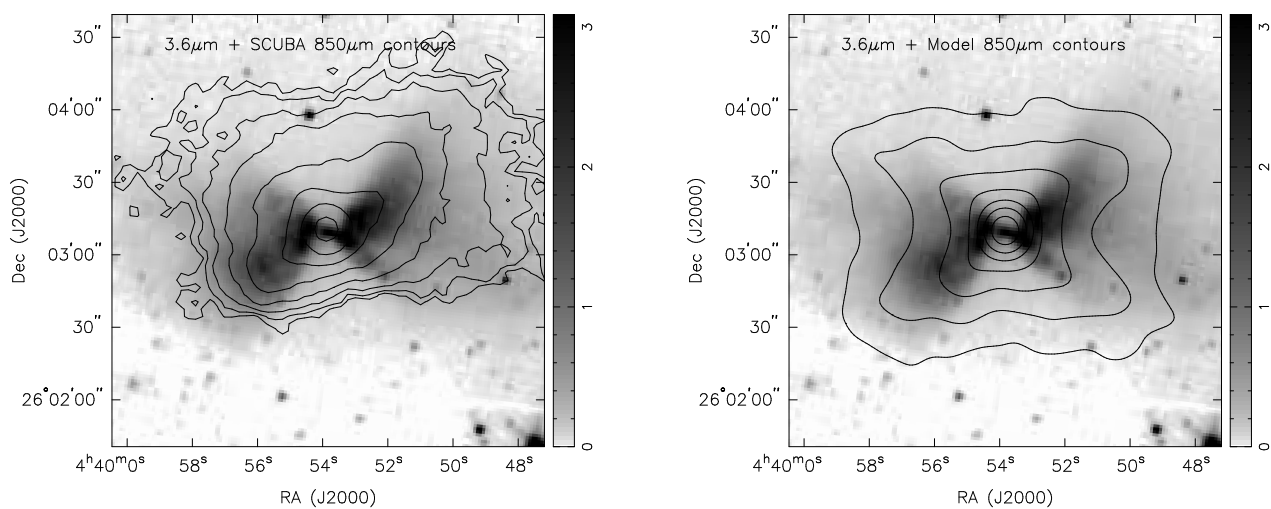}
\end{center}
\caption{Left: IRAC 3.6$\mu$m image with 850$\mu$m SCUBA contours from
\citet{chandler2000}. Right: IRAC 3.6$\mu$m image with model
850$\mu$m contours. Notice the boxiness of the contours and how the south
side flux falls off rapidly as compared to the north side flux in the
SCUBA image. The model image is able to duplicate the boxy features of
the SCUBA data on the inner contours. The SCUBA contours clearly show an
asymmetry corresponding to the bright patches as observed in the IRAC
images. The contours are 2.07, 3.49, 5.58, 9.91, 16.71, 28.16, 47.46,
and 80.0 MJy/sr.}
\label{modelscuba}
\end{figure}

\begin{figure}
\begin{center}
\includegraphics[angle=-90, scale=0.8]{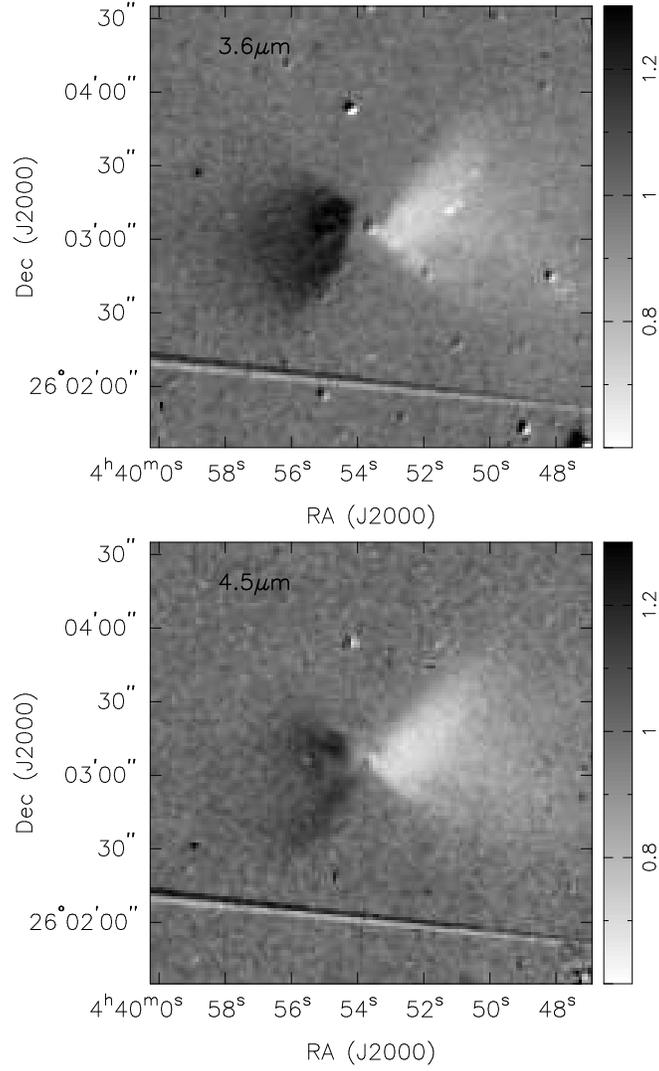}
\end{center}
\epsscale{0.8}
\caption{2004 epoch IRAC images divided by the 2005 epoch IRAC images. The lighter gray to white signifies
less flux in the 2004 epoch, while darker gray means more flux. The variability is present in IRAC Channels
3 and 4, but the noise is much higher. }
\label{2epochs}
\end{figure}

\begin{figure}
\plotone{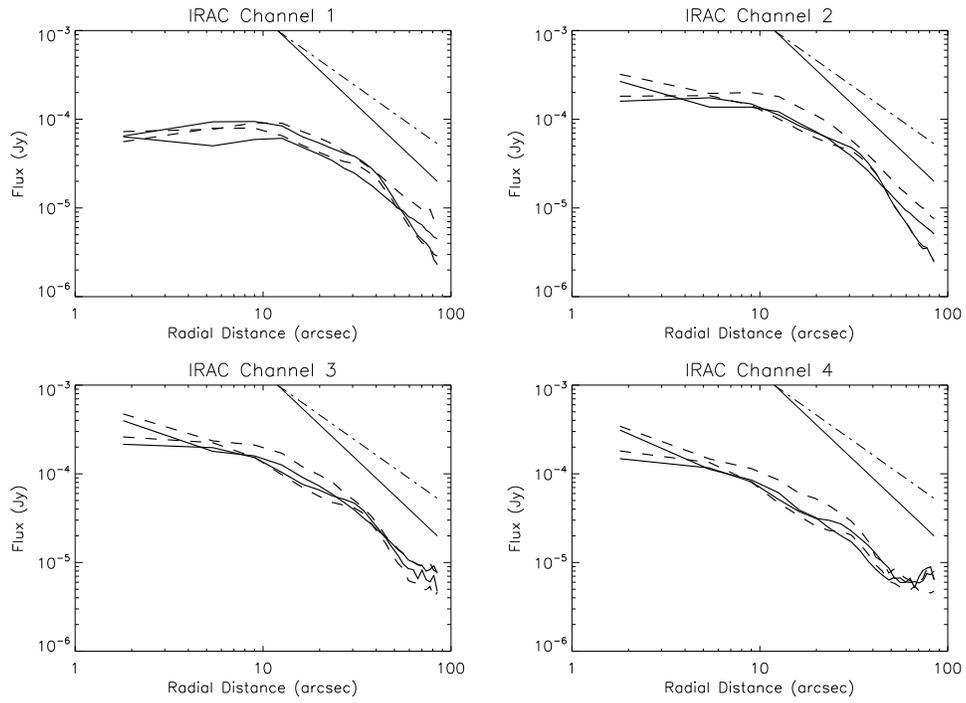}
\epsscale{1.0}
\caption{Average flux in a 3$\farcs$6 width annulus measured radially
from the center. The 2004 (solid lines) and 2005 (dashed lines) epochs
are plotted, there are two lines per epoch corresponding to the east and
west cavities. The straight lines drawn in the upper right correspond
to power-laws of -2 (solid line) and -1.5 (dot-dashed line).}
\label{sectors}
\end{figure}

\begin{figure}

\plotone{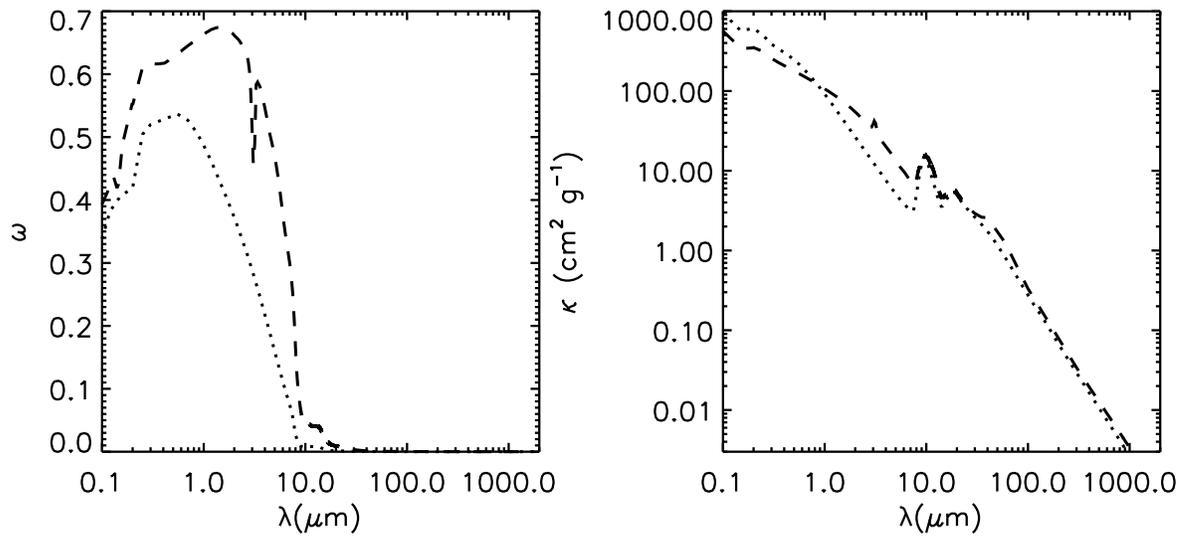}
\epsscale{0.8}

\caption{Albedo (left) and opacity (right) curves of dust
models. The dashed line is the larger grain dust model we used, and the dotted line
is an ISM dust model \citep{kmh1994}.}
\label{dust}
\end{figure}

\begin{figure}

\plotone{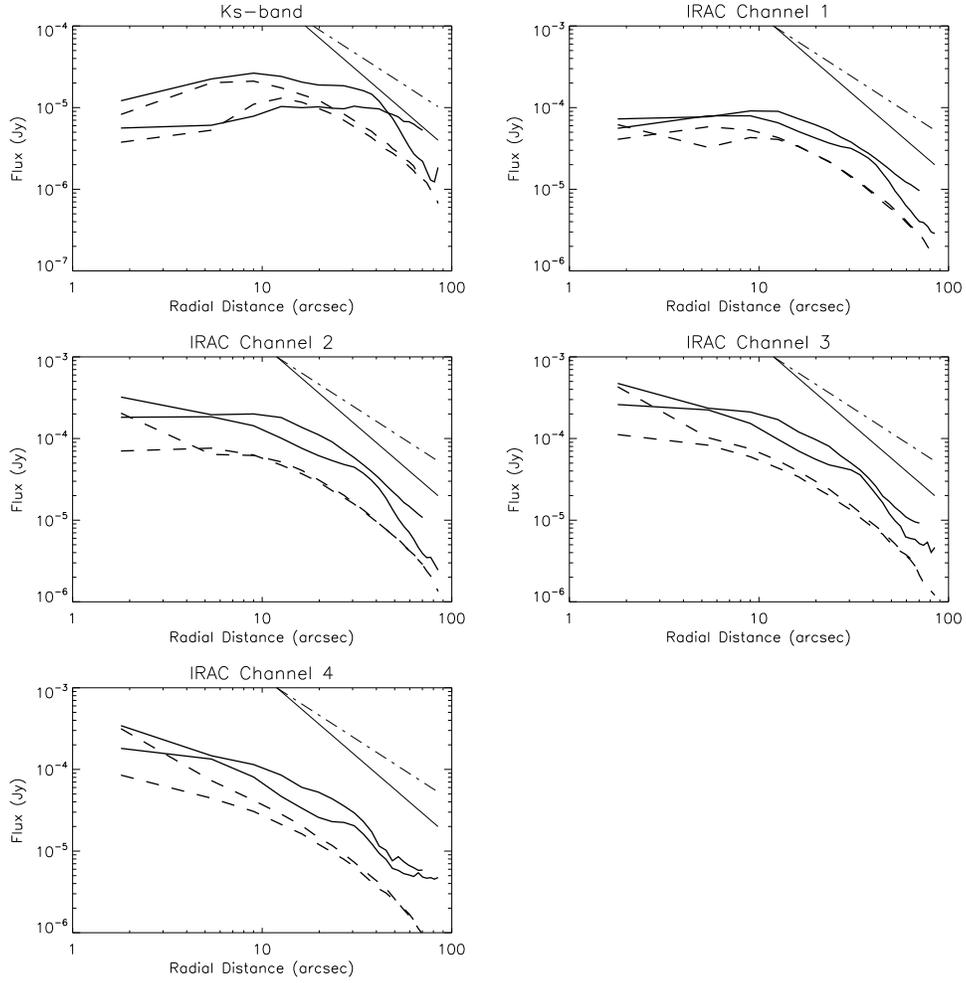}
\epsscale{.8}
\caption{Same as Fig. \ref{sectors} except the data (solid lines) and
model (dashed lines) are plotted. The data the average of the two epochs, except Ks-band.}
\label{modelsectors}
\end{figure}

\begin{figure}

\plotone{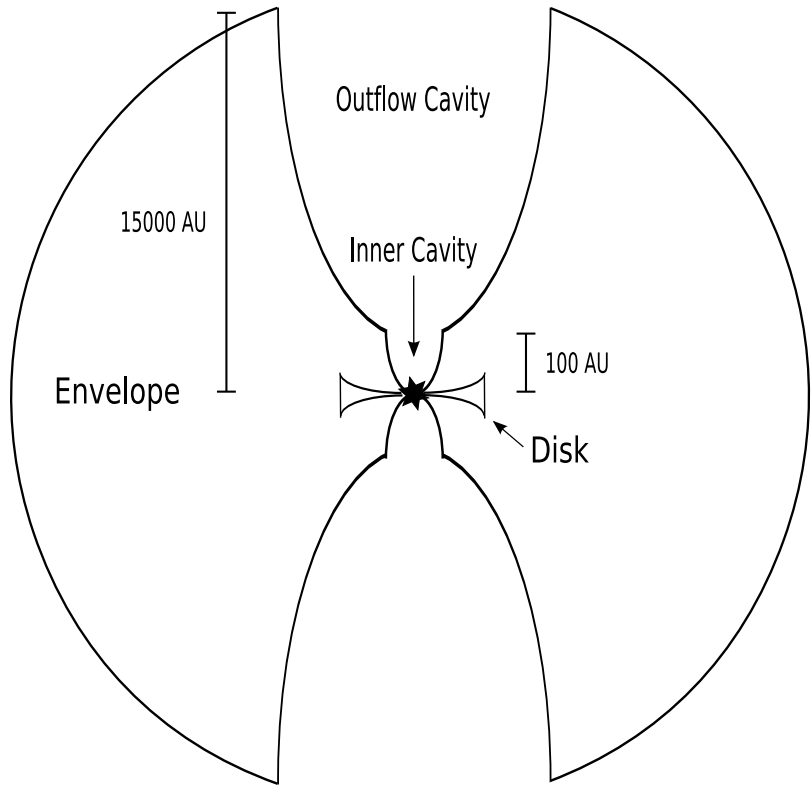}

\epsscale{0.7}
\caption{Schematic representation of modeled cavity and envelope
structure. Inner envelope regions are not drawn to scale.}
\label{cartoon}
\end{figure}

To reproduce the observed `neck' morphology 
we used an inner, narrow outflow cavity with an offset
outer outflow cavity; this `dual-cavity' geometry is shown schematically
in Fig. \ref{cartoon}. In Fig. \ref{zoom}, we compare the results of this model convolved with
the IRAC PSF to the 3.6 $\mu$m IRAC image. Then in the bottom panel of Fig. \ref{zoom},
 we show how the modeled central source actually appears at high resolution.
The bright emission ridges in the IRAC images are produced at the base
of the outer cavity; the inner cavity offsets the outer cavity, producing
neck absorption while still transmitting enough scattered light to form
the central source. The dual-cavity model has some direct support in the image; as shown
in the upper left panel of Fig. \ref{zoom}, the scattered light cavity
appears much narrower near the source, especially on the western side.
Possible physical motivations for the envelope structure we have created
are discussed in \S6.

In order to match the morphology of L1527 using the dual-cavity structure,
the opening angle and cavity shape parameter are relatively well constrained. As
discussed in \S4 the cavities we model are defined by a polynomial of
a variable degree; we refer to the polynomial degree as the shape parameter. The shape parameter determines
how quickly the cavity widens in the inner envelope. The cavity offset is associated with the
inner/outer cavity opening angles and shape
parameters. Our best fitting outer cavity offset is 100 AU,
a larger offset creates a central source that is too resolved. A smaller offset
creates an unresolved central source.

The opening angle of the outflow cavities is defined as the angle measured
from the center of the envelope to the outer envelope radius. Thus, near
the center of the envelope, the cavity opens wider than the opening
angle defined. The opening angle of the outer cavity is modeled to
be 20$^{\circ}$ and the inner cavity opening angle is modeled to be
15$^{\circ}$. The inner cavity is modeled to have a shape parameter b =
1.5 and the outer cavity has a shape parameter b = 1.9. We need to model
the outer cavity to have a shape parameter such that it opens wide a
short distance from the envelope center. The inner cavity then casts
a shadow on the widely opening outer cavity, and creates the apparent
central point source in scattered light. The shape parameters for the 
cavities are empirical. We adjusted the inner and outer shape 
parameters and opening angles until the models appeared similar to the observations.
In the end, the morphology of the scattered light in the modeled cavities is 
similar to the observations.

The scattered light in our model and observations require a large
envelope. To match the scattered light profile in Fig. \ref{modelsectors}
and extent of scattered light in the images,
it was necessary to use an envelope with a 15000 AU outer radius. Larger
radii give too much optical depth to the center of the envelope. With
smaller radii, the scattered light ends prematurely. In addition to
the outer radius, the overall optical depth greatly depends on $R_c$,
$\dot{M}_{env}$, and $M_*$. We assume that the central protostar has a
mass of 0.5\msun, however this is an unknown. The best $\dot{M}_{env}$
that we were able to fit is 1.0 $\times 10^{-5}$ $M_{\sun}$ $yr^{-1}$. KCH93
alternatively express the infall rate as
\begin{equation}
\rho_1 = 5.13 \times 10^{-14}
\left(\frac{\dot{M}_{env}}{10^{-5}M_{\sun}yr^{-1}} \right) \left(
\frac{M_*}{1 M_{\sun}} \right)^{-1/2}.
\end{equation}
For our $\dot{M}_{env}$ and $M_*$, we find $\rho_1$ = 3.75$\times
10^{-14}$ g cm$^{-3}$;  corresponding to a dense circumstellar envelope
appropriate for modeling a Class 0 or I object. A Class 0 object is expected
to have a small $R_c$ and high $\dot{M}_{env}$.

For the outer disk radius we also use the same value as $R_c$, it is
generally accepted that the disk will form inside the centrifugal
radius. Outside $R_c$, the angular momentum is not great enough to
prevent infall. The value of $R_c$ we model must be smaller than the
outer cavity offset in order to create the shadowing effect due to
the control $R_c$ has on the optical depth. The best fitting $R_c$
that we model is 75 AU.

Modeling the inclination of L1527 was trivial as it is observed to be nearly
edge-on. An inclination of 85$^{\circ}$ agrees well with the observations; 
an inclination much lower than 80$^{\circ}$ 
blends the inner cavity scattered light with that of the outer cavity.

\subsection{SED Fitting: Disk Emission}

The fundamental constraint that we need to model is the total
system luminosity.
Integrating over all available photometric data points, we measure a bolometric luminosity (L$_{bol}$)
of 1.97 L$_{\sun}$. The L$_{bol}$ gives an observational constraint on
the total system luminosity that we are modeling. Our measured L$_{bol}$
is consistent with the recent
estimate of 1.9 L$_{\sun}$ by F07. L$_{bol}$ is probably a lower limit
on the total luminosity because the object is observed edge-on and some radiation
from the central protostar may escape through
the outflow cavity without scattering or reprocessing. However, it is unlikely that
the true luminosity is vastly larger because the outflow cavities
do not span a large solid angle as seen from the source; thus most
of the central source emission should be absorbed and reemitted by
the envelope.

The central protostar is chosen to be a 1 L$_{\sun}$ star with an
effective temperature of 4000K, a radius of 2.09 $R_{\sun}$, and a
mass of 0.5\msun. These stellar parameters are chosen from the stellar
birthline of a typical T Tauri star \citep{hartmann1998} as an estimate
for the properties of the central protostar. These parameters are an 
assumption, observations do not constrain the central source.
 Though we know L1527 is probably a binary \citep{loinard2002}, we do not
attempt to model multiple stars and/or disks. Doing so would have
increased complexity and essentially just given us a free parameter as
there are no constraints on the binary pair, other than projected separation.

The brightness of the scattered light combined with an approximate
limit on the bolometric luminosity makes it impossible to explain the
observations without a dominant contribution from the circumstellar
disk at IRAC wavelengths.  For plausible parameters, the infrared
emission of the central protostar is much too faint for a given
total luminosity.  We require a much redder central source SED,
with a high proportion of the luminosity being emitted in the NIR to MIR,
in order to explain the observations. 

For the fixed protostellar parameters, our best results required
an accretion luminosity of $\sim$1.6 L$_{\sun}$ corresponding 
to a disk accretion rate of $3.0 \times 10^{-7}$ M$_{\sun}$ yr$^{-1}$,
so that the true bolometric luminosity in our model is $\sim$2.6 L$_{\sun}$. 
Most of this accretion luminosity is radiated in the optical to ultraviolet by
the accretion shock.  Both the star and the accretion shock heat the
disk, producing the NIR to MIR emission.  This irradiation
heating of the disk dominates the local viscous dissipation and so
the total luminosity, star plus accretion shock, is the important parameter.
Thus in principle we could either make the star fainter and allow the
accretion shock to emit essentially all the radiation; conversely,
we could reduce the disk accretion rate by making the central star brighter,
such that the total luminosity remains constant.  Thus, our model
results in an upper limit to the accretion luminosity and hence
an approximate limit to the disk accretion rate of order $5 \times 10^{-7} M_{\sun}$ yr$^{-1}$.
This disk accretion rate is fairly high compared to most Class II
objects which accrete about $10^{-8}$ $M_{\sun}$ yr$^{-1}$. However, we
cannot distinguish between stellar and accretion luminosity with the model. Though,
significant accretion makes sense in the case of a young protostar.

With the limitations on the bolometric luminosity it was difficult
to produce enough scattered light in the IRAC bands.  Most of the
radiation at these wavelengths comes from the inner disk ``wall'',
i.e. the inner edge of the disk where dust is evaporated
\citep[e.g.][]{dullemond2001,monnier2001,muzerolle2003}.
In our model, this inner disk region emits at about 1400 K, the dust destruction
temperature. The total
luminosity thus depends upon the wall height, and, in our case,
strongly upon the disk wall geometry.
 With our
 assumed stellar and accretion luminosity, the dust destruction radius
 is (14.25R$_{*}$, 0.139 AU). The disk is modeled with a scale height
 h(100AU) = 10.52AU with a flaring power of $\beta$ = 1.125 and a radial
 density exponent $\alpha$ = -2.125. The scale height is calculated using Equation 5.
The total disk thickness at $R_{dd}$ is about 0.12 AU.

\begin{figure}
\begin{center}
\includegraphics[scale=0.7]{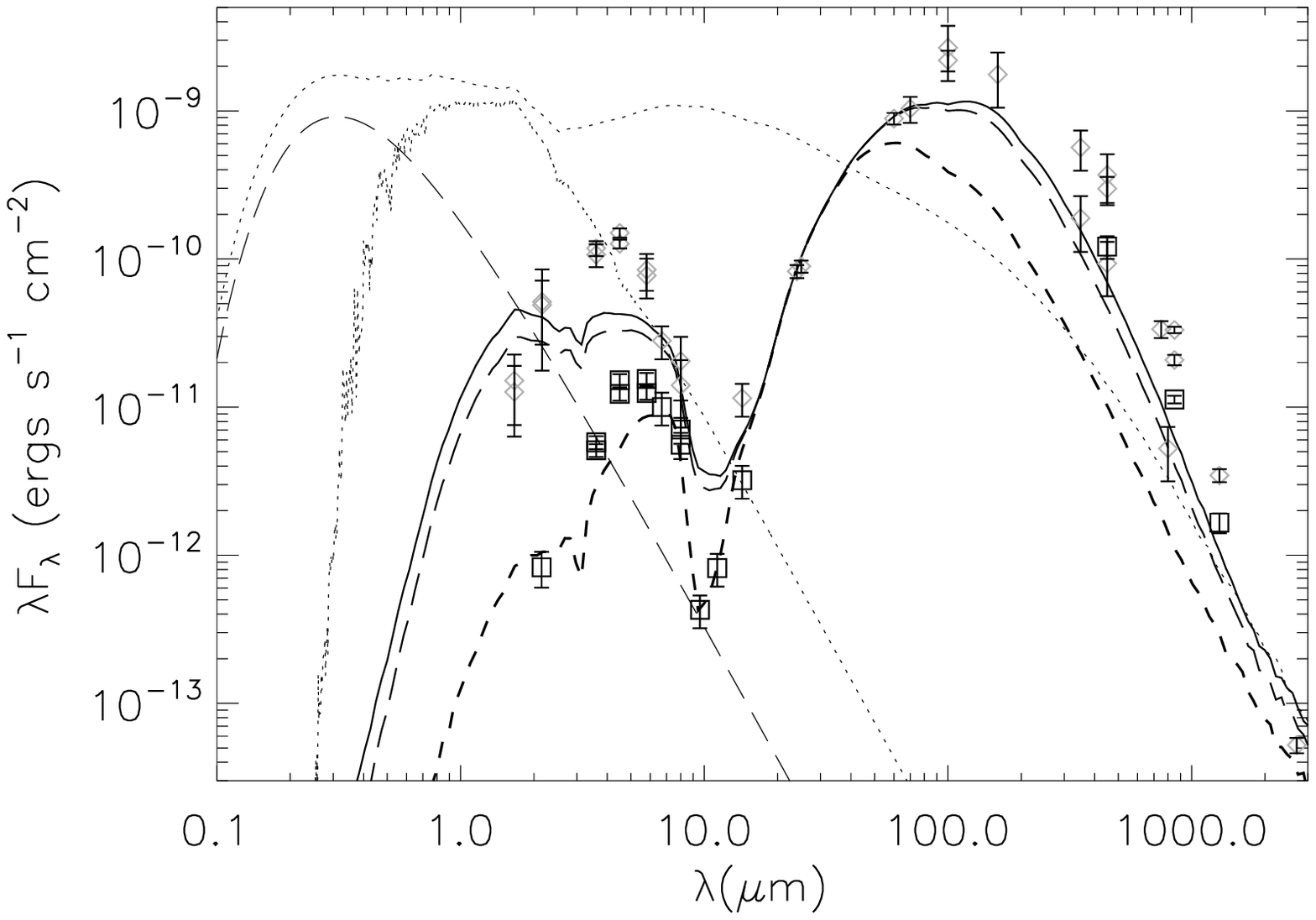}
\includegraphics[scale=0.7]{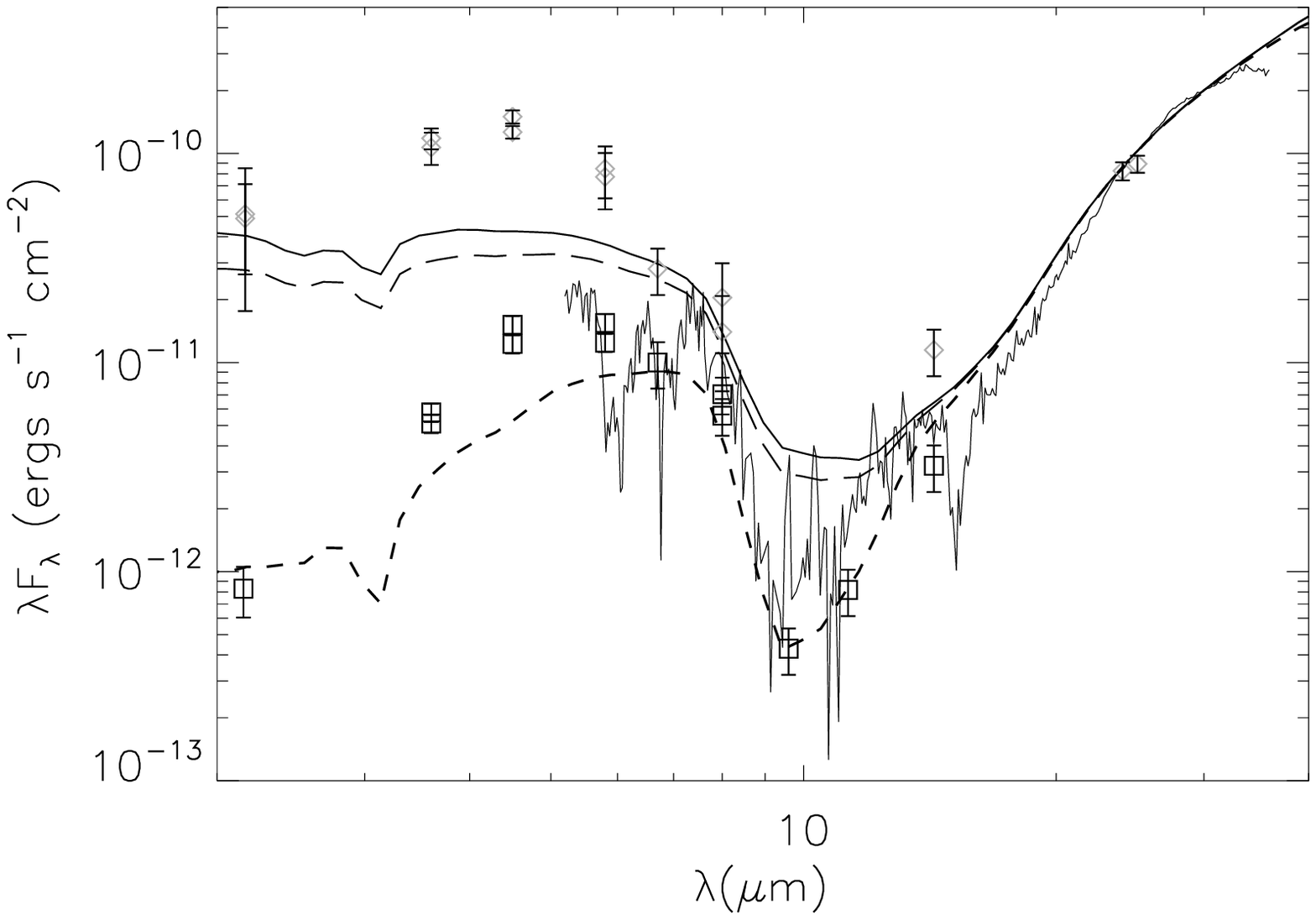}
\end{center}
\caption{The full SED of L1527 (Top) with the model, excluding the IRS spectrum. A zoomed in
portion of the SED (Bottom) is plotted with IRS spectrum from 2 - 40$\mu$m.  
The model SED is plotted for apertures of 10000 AU (solid line), 
6000 AU (long-dashed line), and 1000 AU (short-dashed line).  
Photometry taken with apertures of 71$\farcs$4 (diamonds) and 7$\farcs$14
(boxes) corresponding to 10000 AU and 1000 AU apertures are plotted.
The long wavelength data have varying apertures which are comparable to the 6000 and 10000 AU
apertures. We also plot the spectrum of the central protostar and disk (coarse dotted line) illuminating
the envelope viewed face-on; the 4000K stellar atmosphere (fine dotted line) and the accretion shock 
(long dashed line at short $\lambda$) of the infalling material onto the star. }
\label{sed}
\end{figure}

Even with this bright inner disk wall, we still needed to produce more
NIR to MIR photons in the disk. In particular, our initial
assumption that the disk wall is vertical meant that many wall photons
were not emitted toward the poles, where we need them. We have modified the
inner disk wall to resemble the inner disk structure calculated by
\citet{tann2007}, which creates a wedge-shaped wall resulting from
dust settling \citep{dalessio2006}. The wedge-shaped inner disk wall
will emit optically thick radiation with a temperature distribution as
determined by the radiation transfer. This wedge structure will emit
more NIR and MIR radiation toward the poles, into the outflow cavity,
than the vertical wall. The central star plus disk spectrum as observed
pole-on is shown in Fig. \ref{sed} as the dotted line. The central
star plus disk spectrum has a very strong NIR to MIR excess making
the spectrum very flat from visible to mid infrared wavelengths.

Despite the extraordinary lengths that we have
taken to get more NIR to MIR light out of our model, we still fall
short by about a factor of two at large spatial scales in IRAC channels 1 - 3.  It may be 
that complex structure in the cavity walls, seen
in the real object as bright spots, accounts for the difference. 
An alternate way to increase the scattered light flux is to
include dust in the outflow cavity. This would scatter some of the light
from the central protostar and disk that would otherwise escape. However,
it is not clear whether dust grains would condense in the evacuated
outflow cavity. We experimented with including dust in the outflow cavity and 
found that the shadow cast by the inner cavity is washed out if there is 
dust in the cavity near the central object, where we would expect dust grains 
to possibly condense. 

Even with the difficulties in constructing a model of L1527, our end result 
well approximates the observed morphology in the infrared images (Fig. \ref{model},b), 
IRS spectrum, and SED (Fig. \ref{sed}). Though we did not directly fit the 
submillimeter observations of L1527, our model
image at 850$\mu$m does approximate the observed morphology (Fig. \ref{modelscuba}). Fitting the
observed details in the image are as important as fitting the spectrum itself. 

\begin{figure}
\begin{center}
\includegraphics[scale=0.8]{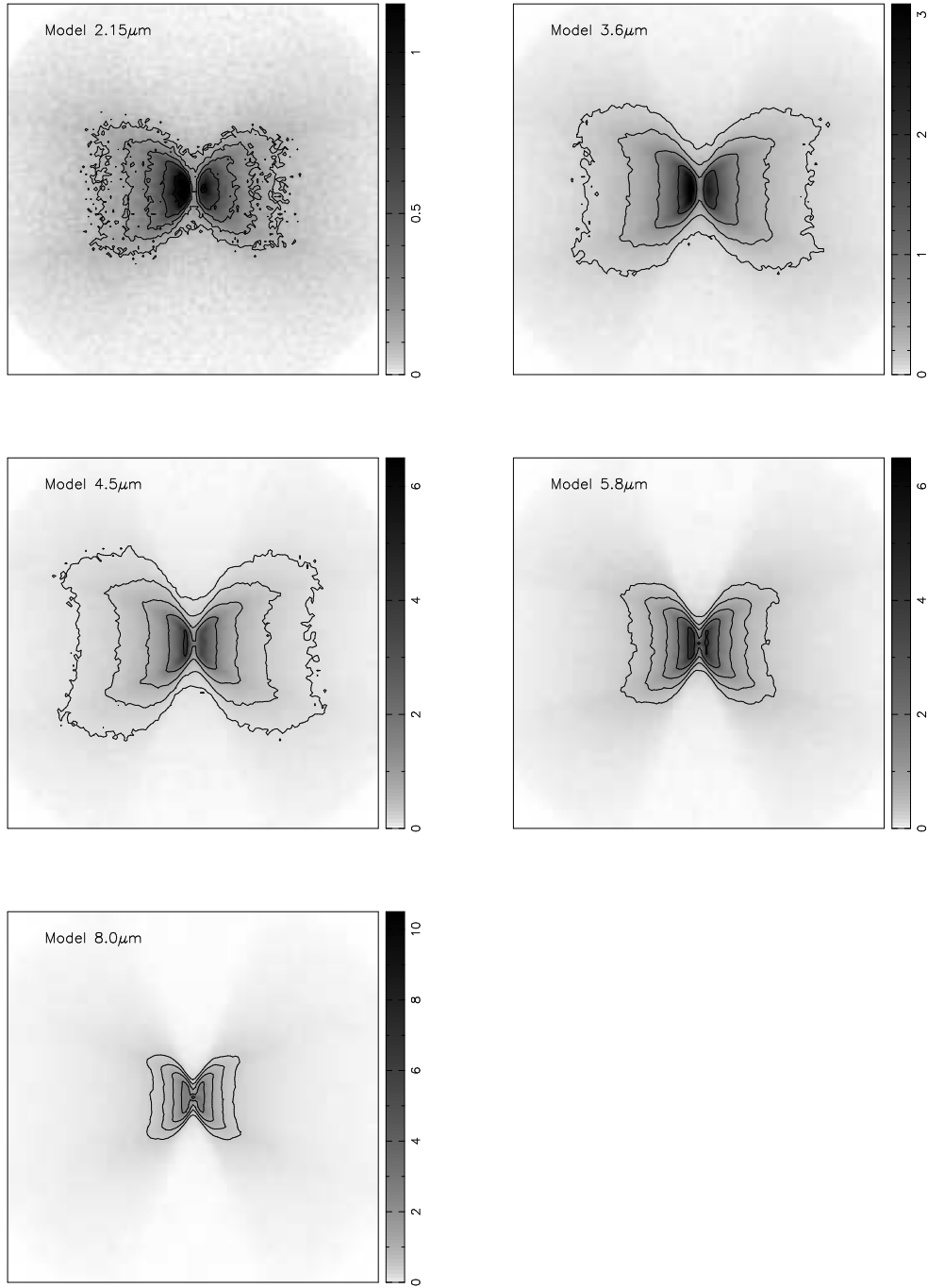}
\end{center}
\caption{Model images generated using a single cavity; all other parameters are 
identical to the dual-cavity model. In comparison to the observations,
these images clearly do a much worse job reproducing the data than the dual-cavity images.
Notice the lack of a central source and thick dark lane in contrast to the dual-cavity model. 
Images are 180$^{\prime\prime}$ on a side, corresponding to $\sim$25000 AU at a
distance of 140pc, units are MJy/sr.}
\label{modelsingle}
\end{figure}

\begin{figure}
\figurenum{13a}
\begin{center}
\includegraphics[scale=.75]{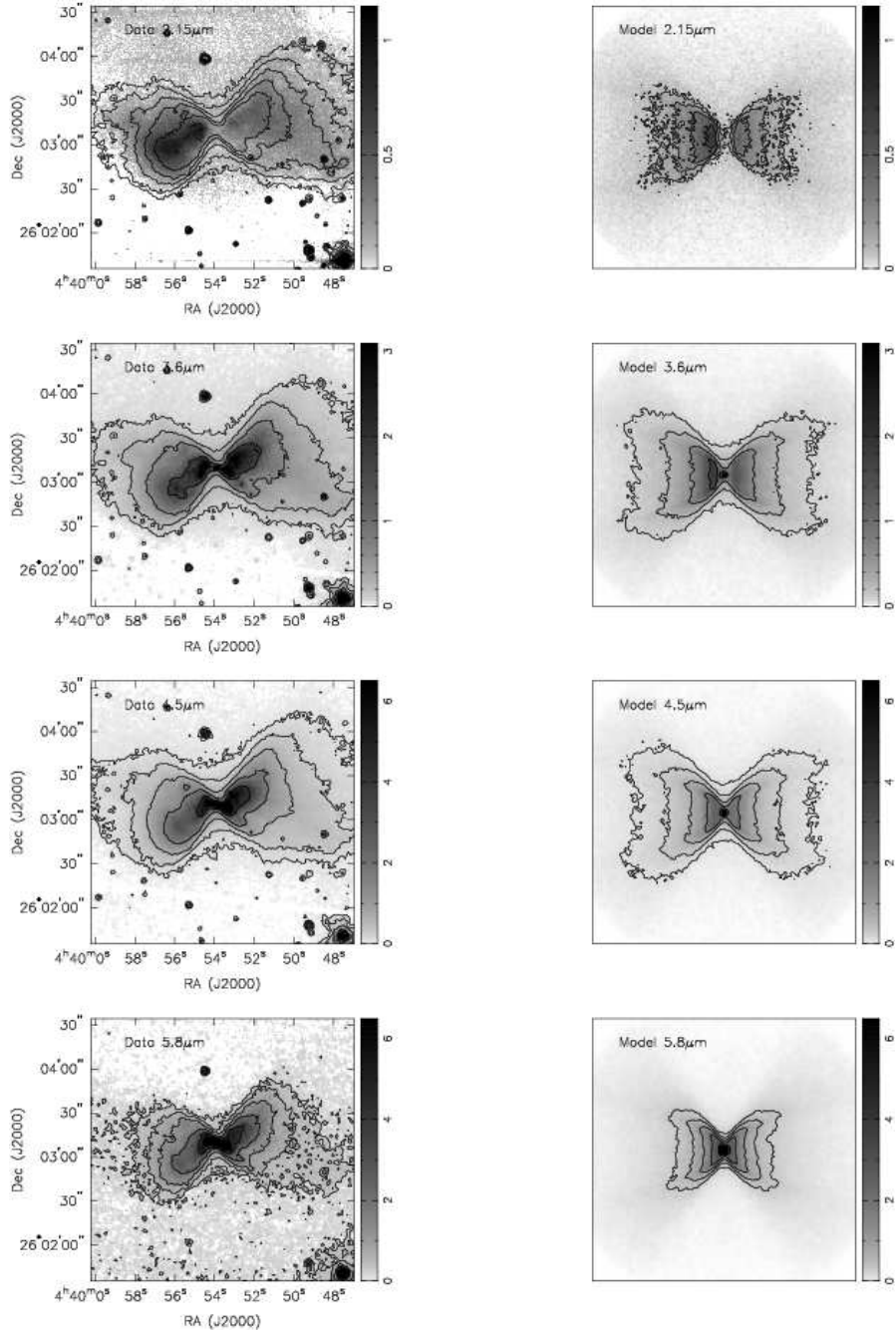}
\end{center}

\caption{Left: L1527 images as in Fig. 2, Right: Best fit model. (a) Images for 2.15, 3.6, 4.5, and 5.8$\mu$m, 
(b) Images for 8.0, 24, and 70$\mu$m. The contours for the L1527
data on the left clearly show the azimuthal asymmetry. Images are 
180$^{\prime\prime}$ on a side, corresponding to $\sim$25000 AU at a
distance of 140pc, units are MJy/sr. }
\label{model}
\end{figure}

Fig. \ref{modelsingle} shows how the L1527 model looks using only a single cavity. Comparison to
Fig. \ref{model},b illustrates the benefits of the dual-cavity model in the case of L1527.
The single-cavity images in Fig. \ref{modelsingle} fail to produce the thick dark lane and
central source between the outflow cavities with the same parameters as the dual-cavity model. 
Thus we have concluded that L1527 is probably not represented by simple bipolar cavities carved
from a TSC envelope.

\section{Discussion}

\subsection{Comparison to Previous Models}

The first attempt at modeling L1527 was done by KCH93. This study used
simple TSC envelopes without cavities to model the spectrum of many Class
0 and Class I sources in Taurus. KCH93 modeled
fixed outer radii of 3000 AU while varying luminosity, envelope density,
centrifugal radius, and inclination. For L1527, KCH93 modeled greatly
larger values of envelope density  $\rho_1$ = 3.16$\times 10^{-13}$
g cm$^{-3}$ compared to our value of $\rho_1$ = 3.75$\times 10^{-14}$ g
cm$^{-3}$ and $R_c$ = 300 AU versus our value of $R_c$ = 75 AU. The high envelope
density modeled by KCH93 is partially explained by the larger $R_c$;
a large $R_c$ decreases the optical depth to the envelope center. The
luminosity of 1.3 L$_{\sun}$ modeled by KCH93 is somewhat lower than our
value of 2.6 L$_{\sun}$. The differences are likely due to the different
modeling techniques and less long wavelength photometry in KCH93. Also, our modeled inclination of
$\sim$85$^{\circ}$ is within their estimated range of 60-90$^{\circ}$;
see Table 2 for a comparison between KCH93 and our modeled parameters. The
differences in our modeled envelope parameters and those of KCH93 may 
be due to the absence of outflow cavities in the KCH93 models as
well as differences in abundances assumed for the dust models.  The large
$R_c$ modeled by KCH93 is necessary to fit the NIR photometry
without including cavities.

F07 also modeled L1527 by SED fitting of multi-wavelength photometry
and the IRS spectrum. This study uses a recent revision of the model
used in KCH93, with outflow cavities. Many of the physical parameters modeled are similar to
our results while some differ. The envelope density, $\rho_1$ =
1.00$\times 10^{-14}$ g cm$^{-3}$ differs from our value of $\rho_1$ =
3.75$\times 10^{-14}$ g cm$^{-3}$. Also, the modeled envelope radius
is similar, 10000 AU compared to our value of 15000 AU. Additionally,
the luminosity modeled by F07 was 1.8 L$_{\sun}$, slightly
less than our value of 2.6 L$_{\sun}$. A limitation of the F07 model
is that there is no way to define an aperture to compare the small
and large scale emission. The model works fine for the 10-50 $\mu$m
range in wavelength if all the flux comes from small scales,
see Fig. \ref{sed}. However, for $\lambda <$ 10$\mu$m and $\lambda >$
100$\mu$m the amount of flux measured can vary greatly depending on
the aperture. Despite the aperture limitations, the model results are
meaningful as the wavelength range of the IRS spectrum is not sensitive to
aperture and the spectrum is fit quite well. See Table 2 for a comparison
of all modeled parameters.

In a recent study employing the \citet{whitney2003a} model, R07 fit
L1527 photometry from IRAC, MIPS, and IRAS against a grid of 200,000
model SEDs. These envelope models are based on the standard bipolar
cavity structure of the \citet{whitney2003a} model as described in \S4. The fits of this model grid seem
to vary depending on the method of fitting. The models which fit SEDs
using the total integrated flux an entire object are consistent with
our results. However, the parameters derived from their resolved source
analysis of L1527 are mostly inconsistent with our results. This analysis
used multiple apertures to examine the integrated flux on small to large
scales. We can account for this inconsistency by observing that their
best fitting models greatly overestimate the long wavelength SED. This
is because the luminosities modeled by R07 are 155 and 20 L$_{\sun}$,
much larger than our modeled luminosity of 2.6 L$_{\sun}$ and the measured L$_{bol}$ of $\sim$2 L$_{\sun}$. The very
high luminosities modeled probably result from the lack of a strong
NIR to MIR excess from the disk and improper weighting of the far-infrared photometry.
The central protostar is modeled to be
the primary source of scattered light photons. In our models, the NIR to MIR 
scattered light comes
from a combination of accretion and the inner disk wall. Also, the R07
analysis did not apply aperture corrections to IRAC channels 3 and
4. Those channels have correction factors less than 1 for apertures
larger than 10$^{\prime\prime}$. In addition to high luminosities, R07
modeled high envelope densities, $\rho_1$ = 2.81 $\times 10^{-12}$ and
8.77 $\times 10^{-13}$ g cm$^{-3}$ compared to our density of $\rho_1$
= 3.75$\times 10^{-14}$ g cm$^{-3}$. The high envelope densities are
necessary to extinct the high stellar luminosity. Also, the lack of an
envelope structure to create the central object in the R07 models makes
it difficult to fit the small and large scale IRAC fluxes simultaneously
with reasonable parameters.

\subsection{Envelope Neck}

To fit the SED  and image of L1527 in multiple
apertures at all wavelengths, we found it necessary to introduce a
new type of outflow cavity structure. We are limited by axisymmetry in the
\citet{whitney2003a} model in ways to create the observed central object
and separated cavities. Our model may be an axisymmetric approximation
to a more complex structure. Having discussed the morphological reasons for adopting the 
`dual-cavity' structure in \S5, we now review the physical motivations
for such a structure. 

\citet{delamarter2000} created MHD simulations of protostellar
outflows originating from a spherical wind driven by the central
protostar. The density structure assumed for the envelope was the
parametrization of the TSC model for the collapse of a flattened sheet
\citep{hartmann1996} interacting with a spherically-symmetric
outflow.  The enhanced density at the equator of the infalling material
resulted in higher ram pressure working against the ram pressure of
the wind, creating a kind of `neck' structure.
Our TSC model does not have an enhanced equatorial density as in
the \citet{delamarter2000} model. However, \citet{ohashi1997} observe a dense,
possibly rotating, inner envelope toroid in L1527 with 
interferometric C$^{18}$O observations. Alternatively, one might suppose that
the ram pressure of the wind varies from the axis in such a way to
produce a `neck', with weaker flow near the equator than the poles;
the effect would be similar because the relevant quantity is the
ratio of wind to inflow ram pressures, not the absolute values.
As outflows are clearly collimated, with high-density jets along
the axis, this supposition is physically plausible. 

It is also worth noting that the `neck' region of L1527
is not axisymmetric.  A precessing jet \citep{gueth1996,ybarra2006} could
partially evacuate the polar regions allowing the spherical outflow to
expand, creating a wider outflow cavity with a standard TSC envelope.

Another possibility is that
binary accretion patterns are responsible for the observed
morphology. Figs. 1 and 2 of \citet{bate1997} show that proto-binary
stars create an evacuated inner region of the protostellar envelope, 
with streams of material falling in at particular longitudes.
 Thus, the `neck' might be related to these streams and evacuated region, while
the outer cavity is dominated by the outflow interacting with a more
axisymmetric envelope.  It is worth considering this further
as  L1527 appears to be a binary system \citep{loinard2002}.

\subsection{Variability and Asymmetry}
In addition to possibly causing asymmetry in the overall structure,
binary interactions could contribute to the variability observed in
L1527. As described in \S3 the variability seems to affect each cavity
lobe independently. Both cavities vary in brightness but one becomes
fainter as the other becomes brighter. The variability is consistent with the scattered
light from a central source; the light travel time out to the
edges of the observed scattered light image is $\sim$0.25 years and
observations were almost a year apart.

Resolved observations of the
HH30 disk have shown substantial variability over the past 15 years
\citep{watson2007}. The mechanism behind these variations is uncertain
but it is conjectured that inner disk warps caused by a companion,
accretion shocks unevenly heating the disk, and/or binary interactions
could be causing uneven patterns of illumination. These proposed mechanisms may 
cause selective illumination of different parts
of the cavity from the disk since most NIR to MIR
photons come from the disk. We may be observing the effects on the envelope of
a process similar to the variability in the disk of HH30.

Variability of the scattered light intensity is not unexpected since we
know Class 0 objects are still accreting material and our models require
high accretion and infall rates. The variability is easier to understand
if the disk accretion luminosity is large, as we have modeled it to be. T Tauri
stars which show the most variation are most commonly accreting systems.

In \S3 we described the azimuthal asymmetry of L1527; the brightness
difference of the north and south sides of the cavities. The radiative
transfer study by \citet{wood2001} showed that misaligned circumstellar
disks could cause asymmetric illumination of the outflow cavities in
the radial and azimuthal directions. In the case of T Tau, Fig. 2 of
\citet{wood2001} shows that one side of the cavity could be significantly
more illuminated than the other, similar to what we observe in L1527. The
contoured IRAC and Ks-band images of L1527 in Fig. \ref{model} clearly show the
azimuthal asymmetry of L1527. Individual circumstellar disks around the
central binary \citep{loinard2002} could be producing the observed brightness asymmetries.

We noted in \S3 that even the submillimeter image shows asymmetry, see Fig. \ref{modelscuba}. 
The boxy contours are extended along the outflow axis, probably due to the heating along the outflow 
cavity walls and possibly dust in the cavity; the contours should be round without a cavity present.
 The contours are more extended
in the southeast and northwest, indicating asymmetric heating in the same locations as 
asymmetric scattered light suggesting that they are related. 

\subsection{Interpretation of Physical Parameters}

In \S5 we described the physical parameters of the model we
have constructed. Two of the most important parameters derived in our
modeling are the mass infall rate $\dot{M}_{env}$ and the mass accretion
rate $\dot{M}_{acc}$. Assuming a central stellar mass of 0.5 $M_{\sun}$, 
we model a $\dot{M}_{acc}$ of 3 $\times$ 10$^{-7}$ $M_{\sun}$ yr$^{-1}$,
an order of magnitude larger than a typical T Tauri star for which 
$\dot{M}_{acc}$ $\sim$ 10$^{-8}$ $M_{\sun}$ yr$^{-1}$. Our high $\dot{M}_{acc}$ modeled is
further supported by the strong forbidden optical emission lines (i.e. [OI], [NII], [SII]) 
and Hydrogen alpha observed by \citet{eiroa1994} and \citet{kenyon1998}; these
observations infer high disk accretion rates as well as jet emission. 

In our model, $\dot{M}_{env}$ is about two orders
of magnitude larger than $\dot{M}_{acc}$, which means that the disk is
rapidly building up mass since it cannot pump material onto the star
quickly enough. Thus the disk will either get very massive in a short
period of time or the excess mass is expelled by the outflow. Also, the widening
of the outflow cavities may be working to quench the mass infall, not
enabling the disk to gain a very large final mass \citep{arce2006}.

With our best fitting $\dot{M}_{env}$, similarly $\rho_1$ which we get
from our assumed stellar mass, we get an envelope with a density similar
to that of many Class I objects as modeled by F07. While
the envelope of L1527 is not substantially more dense than a typical Class
I protostar, the envelope is substantially larger in spatial extent.
 We observe the very extended scattered light with IRAC, but
no other Class I object in Taurus has such extended scattered light except
for L1551 IRS 5 \citep[e.g.][]{osorio2003} and L1551NE \citep{swift2008}.
 However, SED models by F07 fit large
envelopes for most Class I sources for which we do not observe extended
scattered light. This contradiction suggests that the size of the envelope
may not be well constrained by SED modeling and 
spatially resolved observations are required to constrain this parameter.

Related to $\dot{M}_{env}$ is the overall mass in the envelope. The 
mass of our best fitting model was 1.72 M$_{\sun}$. This mass is comparable
to the 2.4 M$_{\sun}$ measured by \citet{benson1989} from ammonia observations. 
Observations of dust emission yield masses of 0.8 M$_{\sun}$ from \citet{shirley2000}, and 
0.9 M$_{\sun}$ from \citet{motte2001}. While these measurements are low, the dust
emissivity may be uncertain by a factor of 3 \citep{shirley2000} which
could account for the difference.

Our best fitting model was not strongly dependent on the value of
$R_c$, but it had to be less than the outer cavity offset to create
the shadow effect. Our disk radius was chosen to match $R_c$ at 75 AU,
given our upper limit on $R_c$ by the cavity offset, our disk will
be small compared to the typical 200 AU size of disks as observed by
\citet{andrews2007}. This implies that disk of L1527 is still being
built up and there are likely complex dynamical interactions on small
scales in the inner envelope as hinted to previously.

\subsection{Class 0 versus I}

The Class 0 definition from \citet{andre1993} requires that the ratio of
submillimeter luminosity to the bolometric luminosity is 5 $\times$ 10$^{-3}$.
 This definition is not exclusive of the I
through III classification of \citet{lada1987}. A Class 0 object can also be a Class I object
if there is detected emission in the NIR to MIR. \textit{Spitzer}
has revealed previously undetected MIR scattered light emission from many
Class 0 protostars \citep{tobin2007, seale2007}.

With the inclusion of outflow cavities in the standard envelope models,
this Class 0 definition immediately becomes muddled. Using L1527 as
an example, with the observed inclination, it is clearly a Class 0
object, all the power is in the long wavelength portion of the SED,
see Fig. \ref{sed}. However, we can predict what our L1527 model would
look like at different inclinations. As L1527 is tilted toward our
line of sight, we see further down the cavity through increasingly
less extinction. The peak of the long wavelength SED does not change,
but more power goes into shorter wavelengths. There is a point
when the object no longer meets the criterion for a Class 0 object and
is a typical Class I, see Fig. \ref{sedinc}. For example, our model
spectra at varying inclinations are similar in shape to the spectra
 of Class I protostars in Taurus from
F07. Specifically, at an inclination of 63$^{\circ}$,
our model is similar to DG Tau B, at 49$^{\circ}$ it is similar to IRAS 04239+2436,
and the same applies for a few others.

From the SEDs shown at various inclinations in Fig. \ref{sedinc}, we show
that L1527 may not necessarily be a Class 0 object and could be a Class I 
if it were oriented differently. Because Class definitions are dependent on the SED,
spatially resolved observations are crucial to our understanding of
protostellar structure and evolution. There are spatially resolved observations of
several Class I protostars in Taurus from \citet{padgett1999}; 
these Class I objects are quite different from L1527. The scattered
light morphologies observed by \citet{padgett1999} are all very small in
spatial extent ($\sim$2000 AU) compared to L1527, which is extended over
nearly 30000 AU. Without resolved observations, we would be ignorant of
their inherent differences.

\subsection{General Applicability of the Model}
An important question must be asked of our results; is L1527 a special
case in requiring a modified cavity structure to fit its morphology or
are our results more generally applicable? Some
protostars associated with scattered light nebulae seem to show a
similar morphology to L1527; a point source located between two outflow
cavities. However, not all sources have cavity structures as bright as
L1527 and not all objects appear to have a cavity separation like L1527,
although this may be an issue of angular resolution and inclination.

The observations by \citet{padgett1999} appear to be consistent with a
single outflow cavity model; \citet{stark2006} modeled these objects
as such. However, in contrast to L1527,
these objects seem to be more evolved, Class I objects. Protostars may evolve
such that a more simplified model will suffice as the infall phase is
ending. The dual-cavity model may only be applicable to the very young,
high mass infall phase.

\citet{seale2007} cataloged many remarkably bright outflow cavity
structures in archived observations from the \textit{Spitzer Space
Telescope}. Several objects (IRAS 05491+0247, IRAS 18148-0440, L1251B,
and others) show bright cavity structures with an apparent point source
in the middle. From this catalog of scattered light nebulae observed
with IRAC, it appears that the model we have developed for L1527 may be
more than just a special case.

We stress that this morphology of an observed central object and separated
outflow cavities for Class 0 sources is only observed between 3$\mu$m
and 8$\mu$m thus far. Shorter wavelengths are too extincted to detect the
central object and longer wavelengths wash out the separation because the
dust is less opaque to the incident radiation. Objects with a less dense
circumstellar envelope may show a central object at K-band as seen in
NICMOS observations by \citet{meakin2003} and \citet{hartmann1999}. 
Also, an object must be viewed
nearly edge-on for this morphology to be apparent. Sources inclined much
less than 80$^{\circ}$ will blend the central object with the blue-shifted
outflow cavity. Though, a gap caused by the inner envelope shadowing
should still be visible between the red and blue-shifted cavities.

In addition to the necessity of edge-on observation, objects also must
be nearby. The central object of L1527 is only separated from the bright
outflow cavity by $\sim$3$\farcs$6. If the distance to L1527 were twice
what it is, the separation may be smeared out by the PSF. These criteria
for observation of this morphology will create a selection effect;
however, increased spatial resolution in the NIR to MIR is not far away
and detailed modeling may be able to disentangle the inclination effects.

\subsection{Dust Properties and Emission Processes}

As discussed in \S5, we were not able to completely model the SED
with the assumed dust model. Our dust model is not unique, there could
be different dust grain size distributions in different parts of the
envelope; large grains in the dense inner regions and small grains in the outer
regions. \textit{Spitzer} has provided a wealth of data with which more precise
dust models may be developed; most importantly, the infrared extinction
law in the IRAC wavelengths has been proven to be flatter than expected
\citep{indebetouw2005, flah2007, zuninga2007}. This flat extinction law
implies that grain growth has occurred in dense regions \citep{zuninga2007}.
Thus protostars
start out with dust that is larger than general ISM dust. These extinction law
measurements support our adoption of a larger dust grain model for our modeled
envelope. The larger dust grains have significantly more albedo in the 
NIR to MIR.

Also, as shown in Fig. \ref{sed}, the model spectrum tends to fall below most
of the submillimeter data points.  This trend again indicates that
there are details missing from our assumed dust model.
Other, more complex
organics may more accurately represent an envelope dust model
\citep[e.g.][]{pollack1994}. Also, heating of the outer envelope from
the interstellar radiation field could be responsible for the observed
submillimeter fluxes slightly disagreeing with the model. The upcoming
\textit{Herschel Space Telescope} will give new insights into the submillimeter
properties of dust.

In addition to the scattered light emission, there is a possibility of some 
shock-excited molecular line emission resulting from the jets and outflows in the cavities.
The strong Herbig-Haro spectrum detected in the cavity of L1527 \citep{eiroa1994, kenyon1998}
suggests that the outflow cavities could have some NIR to MIR emission from H$_2$
ro-vibrational lines. Many H$_2$ lines are present throughout all the
IRAC bands. However, IRS observations of protostars in F07
showed no evidence for strong emission lines at IRAC wavelengths. Our narrowband H$_2$ (2.12$\mu$m)
observations of L1527 did detect some weak emission in excess of observations with the 
H$_2$ (2.09$\mu$m) continuum filter. However, the H$_2$ emission is at most 8\% of the 
total Ks-band flux which is within our errors for the flux at Ks-band and would not greatly
affect modeling. 

In studies of other protostars, \citet{fuller1995}
detected H$_2$ emission in the cavity of L483. However, the emission was localized
to a small section of the cavity, appearing to be an outflow knot within the cavity.
Also, \citet{hartmann1999} did not detect any excess H$_2$ emission from the IRAS 04325+2402 
scattered light nebula. Therefore, the dominant emission mechanism in the NIR to MIR
for protostars is scattered continuum emission from the central star and disk and 
line emission should not significantly contribute to the total flux.

\subsection{Predictions}
Our dual-cavity model can probably be tested easily by high resolution
imaging of the L1527 central object in the MIR. A model image
with 0$\farcs$11 pixels convolved with 0$\farcs$5 seeing is shown in 
Fig. \ref{simobs}.
These model images simulate what could be observed from a ground-based
MIR telescope at L$^{\prime}$, M, or even 7.7$\mu$m.
Imaging of this structure is within the
capabilities of even modest aperture, infrared optimized telescopes,
so long as the PSF is well sampled. As discussed in \S3, the central object appears
marginally resolved in the IRAC data, but the resolution is too low
to observe the small scale structures we are modeling. We would probably see some of the inner
envelope structure in the IRAC images if the PSF was more finely sampled at 3.6$\mu$m.

Excellent seeing or adaptive optics (AO) are required to observe the inner envelope structure.
But AO observations are difficult for protostars due to the high extinctions
obscuring guide stars near the central object. However, at 7.7$\mu$m the seeing typically can be better
than 0$\farcs$5 naturally, thus imaging farther into the MIR may be ideal for 
observing the inner envelope.

Our model may be an axisymmetric approximation to some complicated feature,
thus high resolution observations may not completely reflect the image shown
in Fig. \ref{simobs}.  We expect
high resolution imaging to show some resolved structure that is probably
responsible for the complicated morphology observed.

\section{Conclusions}

We have presented a detailed analysis of infrared scattered light 
observations of the Class 0 object L1527 (IRAS 04368+2557). L1527 presents itself as a very intriguing
object, having bright scattered light emission at NIR to MIR wavelengths,
and outflow cavities which are asymmetric in brightness and shape. The
extended emission of L1527 is also variable; observations taken a year
apart show significant variability. This variability is not constant for
the entire object, the two cavities seem to brighten and fade independently.

We have constructed a model of L1527 which approximates the observed
morphology of L1527 in the IRAC and Ks-band images as well as the overall SED. The model
uses a modified cavity structure with a narrow cavity near the central
protostar and a wider cavity offset from the central protostar. This
modified cavity structure creates an apparent central point source in
the IRAC images from light scattering on the inner cavity. The inner
cavity also casts a shadow on the outer cavity creating the observed
separation and dark lane between the outflow cavities. While modeling is
inherently not unique at defining the structure of objects, modeling can tell us what 
the structure is not. This paper shows that L1527 is probably not represented
by simple bipolar cavities carved from a TSC envelope, but possibly represented by the structure we have
created.

We have demonstrated the importance of simultaneously fitting the
morphology of the spatially resolved images as well as the SED. Only fitting the SED
could lead to modeling unphysical parameters if the parameter space is
not adequately restricted. It can be argued that we also have essentially
turned knobs until the imaged looked right.  Though, there are clearly asymmetries that are
not reflected in our model which impact our SED and image fit. 
Within the limits of axisymmetry our model fits the observed images and
SED quite well. Also, the robustness of our SED fit is probably limited
by the uncertain envelope dust model. Our modeled cavity
structure may be an axisymmetric approximation of a complex structure
resulting from outflow-envelope interactions or binary accretion effects.

The dual-cavity model we have developed is specifically for L1527, however,
the model may be more generally applicable. Archived observations from the
\textit{Spitzer Space Telescope} of other Class 0 protostars show similar
features to L1527; a central object observed between the outflow cavities
\citep{seale2007}. This central object can probably be resolved using
current ground-based MIR instrumentation.  This paper demonstrates that observations of the
circumstellar envelope by scattered light in conjunction with radiative
transfer modeling provide new insights into envelope structure which
were previously beyond our reach. 

\acknowledgements
The authors wish to thank B. A. Whitney making her radiative transfer
model public and helpful discussions regarding its use and
C. Chandler for providing the SCUBA images. We thank the anonymous
referee for helpful comments which improved the final paper. We also thank
L. W. Looney for his early guidance and Z. Zhu for his assistance
in taking the L1527 observations with TIFKAM. We are grateful to the staff
of MDM Observatory for their support during TIFKAM observations.
TIFKAM was funded by The Ohio State University, 
the MDM consortium, MIT, and NSF grant AST-9605012. The HAWAII2 array used 
in TIFKAM was purchased with an NSF Grant to Dartmouth University.
 This publication
makes use of data products from the Two Micron All Sky Survey, which
is a joint project of the University of Massachusetts and the Infrared
Processing and Analysis Center/California Institute of Technology,
funded by the National Aeronautics and Space Administration and the
National Science Foundation.  
J.T., L.H., and N.C. acknowledge support from the University of Michigan.

\bibliographystyle{apj}
\bibliography{ms.bbl}

\begin{thebibliography}{63}
\expandafter\ifx\csname natexlab\endcsname\relax\def\natexlab#1{#1}\fi

\bibitem[Adams et al.(1987)]{adams1987} Adams, F.~C., Lada, 
C.~J., \& Shu, F.~H.\ 1987, \apj, 312, 788

\bibitem[{{Andre} {et~al.}(1993){Andre}, {Ward-Thompson}, \&
  {Barsony}}]{andre1993}
{Andre}, P., {Ward-Thompson}, D., \& {Barsony}, M. 1993, \apj, 406, 122

\bibitem[{{Andrews} \& {Williams}(2007)}]{andrews2007}
{Andrews}, S.~M., \& {Williams}, J.~P. 2007, \apj, 659, 705

\bibitem[{{Arce} \& {Sargent}(2006)}]{arce2006}
{Arce}, H.~G., \& {Sargent}, A.~I. 2006, \apj, 646, 1070

\bibitem[{{Bate} \& {Bonnell}(1997)}]{bate1997}
{Bate}, M.~R., \& {Bonnell}, I.~A. 1997, \mnras, 285, 33

\bibitem[{{Beichman} {et~al.}(1988){Beichman}, {Neugebauer}, {Habing}, {Clegg},
  \& {Chester}}]{iras}
{Beichman}, C.~A., {Neugebauer}, G., {Habing}, H.~J., {Clegg}, P.~E., \&
  {Chester}, T.~J., eds. 1988, {Infrared astronomical satellite (IRAS) catalogs
  and atlases. Volume 1: Explanatory supplement}, Vol.~1

\bibitem[{{Benson} \& {Myers}(1989)}]{benson1989}
{Benson}, P.~J., \& {Myers}, P.~C. 1989, \apjs, 71, 89

\bibitem[{{Bergin} \& {Tafalla}(2007)}]{bergin2007}
{Bergin}, E.~A., \& {Tafalla}, M. 2007, ArXiv e-prints, 705

\bibitem[{{Bontemps} {et~al.}(1996){Bontemps}, {Andre}, {Terebey}, \&
  {Cabrit}}]{bontemps1996}
{Bontemps}, S., {Andre}, P., {Terebey}, S., \& {Cabrit}, S. 1996, \aap, 311,
  858

\bibitem[{{Calvet} \& {Gullbring}(1998)}]{calvet1998}
{Calvet}, N., \& {Gullbring}, E. 1998, \apj, 509, 802

\bibitem[Calvet et al.(1994)]{calvet1994} Calvet, N., Hartmann, 
L., Kenyon, S.~J., \& Whitney, B.~A.\ 1994, \apj, 434, 330 

\bibitem[{{Chandler} \& {Richer}(2000)}]{chandler2000}
{Chandler}, C.~J., \& {Richer}, J.~S. 2000, \apj, 530, 851

\bibitem[Chen et al.(1995)]{chen1995} Chen, H., Myers, P.~C., 
Ladd, E.~F., \& Wood, D.~O.~S.\ 1995, \apj, 445, 377 

\bibitem[{{Cunningham} {et~al.}(2005){Cunningham}, {Frank}, \&
  {Hartmann}}]{cunningham2005}
{Cunningham}, A., {Frank}, A., \& {Hartmann}, L. 2005, \apj, 631, 1010

\bibitem[D'Alessio et al.(2001)]{dalessio2001} D'Alessio, P., 
Calvet, N., \& Hartmann, L.\ 2001, \apj, 553, 321 

\bibitem[{{D'Alessio} {et~al.}(2006){D'Alessio}, {Calvet}, {Hartmann},
  {Franco-Hern{\'a}ndez}, \& {Serv{\'{\i}}n}}]{dalessio2006}
{D'Alessio}, P., {Calvet}, N., {Hartmann}, L., {Franco-Hern{\'a}ndez}, R., \&
  {Serv{\'{\i}}n}, H. 2006, \apj, 638, 314

\bibitem[{{Delamarter} {et~al.}(2000){Delamarter}, {Frank}, \&
  {Hartmann}}]{delamarter2000}
{Delamarter}, G., {Frank}, A., \& {Hartmann}, L. 2000, \apj, 530, 923

\bibitem[Draine \& Lee(1984)]{draine1984} Draine, B.~T., \& Lee, 
H.~M.\ 1984, \apj, 285, 89 

\bibitem[{{Dullemond} {et~al.}(2001){Dullemond}, {Dominik}, \&
  {Natta}}]{dullemond2001}
{Dullemond}, C.~P., {Dominik}, C., \& {Natta}, A. 2001, \apj, 560, 957

\bibitem[{{Eiroa} {et~al.}(1994){Eiroa}, {Miranda}, {Anglada}, {Estalella}, \&
  {Torrelles}}]{eiroa1994}
{Eiroa}, C., {Miranda}, L.~F., {Anglada}, G., {Estalella}, R., \& {Torrelles},
  J.~M. 1994, \aap, 283, 973

\bibitem[{{Fazio} {et~al.}(2004){Fazio}, {Hora}, {Allen}, {Ashby}, {Barmby},
  {Deutsch}, {Huang}, {Kleiner}, {Marengo}, {Megeath}, {Melnick}, {Pahre},
  {Patten}, {Polizotti}, {Smith}, {Taylor}, {Wang}, {Willner}, {Hoffmann},
  {Pipher}, {Forrest}, {McMurty}, {McCreight}, {McKelvey}, {McMurray}, {Koch},
  {Moseley}, {Arendt}, {Mentzell}, {Marx}, {Losch}, {Mayman}, {Eichhorn},
  {Krebs}, {Jhabvala}, {Gezari}, {Fixsen}, {Flores}, {Shakoorzadeh}, {Jungo},
  {Hakun}, {Workman}, {Karpati}, {Kichak}, {Whitley}, {Mann}, {Tollestrup},
  {Eisenhardt}, {Stern}, {Gorjian}, {Bhattacharya}, {Carey}, {Nelson},
  {Glaccum}, {Lacy}, {Lowrance}, {Laine}, {Reach}, {Stauffer}, {Surace},
  {Wilson}, {Wright}, {Hoffman}, {Domingo}, \& {Cohen}}]{fazio2004}
{Fazio}, G.~G., et al.  2004, \apjs, 154, 10

\bibitem[{{Flaherty} {et~al.}(2007)}]{flah2007}
{Flaherty} et al. 2007,
  \apj, 663, 1069

\bibitem[{{Fuller} {et~al.}(1995){Fuller}, {Lada}, {Masson}, \&
  {Myers}}]{fuller1995}
{Fuller}, G.~A., {Lada}, E.~A., {Masson}, C.~R., \& {Myers}, P.~C. 1995, \apj,
  453, 754

\bibitem[{{Furlan} {et~al.}(2007)}]{furlan2007}
{Furlan}, E., et al. 2007, ApJ in press.

\bibitem[{{Gueth} {et~al.}(1996){Gueth}, {Guilloteau}, \&
  {Bachiller}}]{gueth1996}
{Gueth}, F., {Guilloteau}, S., \& {Bachiller}, R. 1996, \aap, 307, 891

\bibitem[{{Hartmann}(1998)}]{hartmann1998}
{Hartmann}, L. 1998, {Accretion Processes in Star Formation} (Accretion
  processes in star formation / Lee Hartmann.~Cambridge, UK ; New York :
  Cambridge University Press, 1998.~(Cambridge astrophysics series ; 32) ISBN
  0521435072.)

\bibitem[{{Hartmann} {et~al.}(1996){Hartmann}, {Calvet}, \&
  {Boss}}]{hartmann1996}
{Hartmann}, L., {Calvet}, N., \& {Boss}, A. 1996, \apj, 464, 387

\bibitem[Hartmann et al.(1999)]{hartmann1999} Hartmann, L., Calvet, 
N., Allen, L., Chen, H., \& Jayawardhana, R.\ 1999, \aj, 118, 1784 

\bibitem[{{Hartmann} {et~al.}(2005){Hartmann}, {Megeath}, {Allen}, {Luhman},
  {Calvet}, {D'Alessio}, {Franco-Hernandez}, \& {Fazio}}]{hartmann2005}
{Hartmann}, L., et al. 2005, \apj, 629, 881


\bibitem[Hogerheijde et al.(1998)]{hoger1998} Hogerheijde, M.~R., 
van Dishoeck, E.~F., Blake, G.~A., 
\& van Langevelde, H.~J.\ 1998, \apj, 502, 315 

\bibitem[{{Houck} {et~al.}(2004){Houck}, {Roellig}, {van Cleve}, {Forrest},
  {Herter}, {Lawrence}, {Matthews}, {Reitsema}, {Soifer}, {Watson}, {Weedman},
  {Huisjen}, {Troeltzsch}, {Barry}, {Bernard-Salas}, {Blacken}, {Brandl},
  {Charmandaris}, {Devost}, {Gull}, {Hall}, {Henderson}, {Higdon}, {Pirger},
  {Schoenwald}, {Sloan}, {Uchida}, {Appleton}, {Armus}, {Burgdorf},
  {Fajardo-Acosta}, {Grillmair}, {Ingalls}, {Morris}, \& {Teplitz}}]{irs2004}
{Houck}, J.~R., et al. 2004, \apjs, 154, 18

\bibitem[Hunt et al.(1998)]{hunt1998} Hunt, L.~K., Mannucci, F., 
Testi, L., Migliorini, S., Stanga, R.~M., Baffa, C., Lisi, F., \& Vanzi, 
L.\ 1998, \aj, 115, 2594 

\bibitem[{{Indebetouw} {et~al.}(2005){Indebetouw}, {Mathis}, {Babler}, {Meade},
  {Watson}, {Whitney}, {Wolff}, {Wolfire}, {Cohen}, {Bania}, {Benjamin},
  {Clemens}, {Dickey}, {Jackson}, {Kobulnicky}, {Marston}, {Mercer},
  {Stauffer}, {Stolovy}, \& {Churchwell}}]{indebetouw2005}
{Indebetouw}, R., et al. 2005, \apj, 619, 931

\bibitem[{{Kenyon} {et~al.}(1998){Kenyon}, {Brown}, {Tout}, \&
  {Berlind}}]{kenyon1998}
{Kenyon}, S.~J., {Brown}, D.~I., {Tout}, C.~A., \& {Berlind}, P. 1998, \aj,
  115, 2491

\bibitem[{{Kenyon} {et~al.}(1993){Kenyon}, {Calvet}, \& {Hartmann}}]{kch1993}
{Kenyon}, S.~J., {Calvet}, N., \& {Hartmann}, L. 1993, \apj, 414, 676

\bibitem[{{Kim} {et~al.}(1994){Kim}, {Martin}, \& {Hendry}}]{kmh1994}
{Kim}, S.-H., {Martin}, P.~G., \& {Hendry}, P.~D. 1994, \apj, 422, 164

\bibitem[Lada (1987)]{lada1987} Lada, C.~J.\ 1987, Star Forming 
Regions, 115, 1 

\bibitem[{{Ladd} {et~al.}(1991){Ladd}, {Adams}, {Fuller}, {Myers}, {Casey},
  {Davidson}, {Harper}, \& {Padman}}]{ladd1991}
{Ladd}, E.~F., et al. 1991, \apj, 382, 555

\bibitem[{{Li} \& {Draine}(2001)}]{draine2001}
{Li}, A., \& {Draine}, B.~T. 2001, \apj, 554, 778

\bibitem[{{Loinard} {et~al.}(2002){Loinard}, {Rodr{\'{\i}}guez}, {D'Alessio},
  {Wilner}, \& {Ho}}]{loinard2002}
{Loinard}, L., {Rodr{\'{\i}}guez}, L.~F., {D'Alessio}, P., {Wilner}, D.~J., \&
  {Ho}, P.~T.~P. 2002, \apjl, 581, L109

\bibitem[Looney et al.(2007)]{looney2008} Looney, L.~W., Tobin, 
J.~J., \& Kwon, W.\ 2007, \apjl, 670, L131 

\bibitem[{{Meakin} {et~al.}(2003){Meakin}, {Bieging}, {Latter}, {Hora}, \&
  {Tielens}}]{meakin2003}
{Meakin}, C.~A., {Bieging}, J.~H., {Latter}, W.~B., {Hora}, J.~L., \&
  {Tielens}, A.~G.~G.~M. 2003, \apj, 585, 482

\bibitem[{{Motte} \& {Andr{\'e}}(2001)}]{motte2001}
{Motte}, F., \& {Andr{\'e}}, P. 2001, \aap, 365, 440

\bibitem[{{Muzerolle} {et~al.}(2003){Muzerolle}, {Calvet}, {Hartmann}, \&
  {D'Alessio}}]{muzerolle2003}
{Muzerolle}, J., {Calvet}, N., {Hartmann}, L., \& {D'Alessio}, P. 2003, \apjl,
  597, L149

\bibitem[{{Noriega-Crespo} {et~al.}(2004){Noriega-Crespo}, {Morris}, {Marleau},
  {Carey}, {Boogert}, {van Dishoeck}, {Evans}, {Keene}, {Muzerolle},
  {Stapelfeldt}, {Pontoppidan}, {Lowrance}, {Allen}, \& {Bourke}}]{noriega2004}
{Noriega-Crespo}, A., et al. 2004, \apjs, 154, 352

\bibitem[{{Ohashi} {et~al.}(1997){Ohashi}, {Hayashi}, {Ho}, \&
  {Momose}}]{ohashi1997}
{Ohashi}, N., {Hayashi}, M., {Ho}, P.~T.~P., \& {Momose}, M. 1997, \apj, 475,
  211

\bibitem[Osorio et al.(2003)]{osorio2003} Osorio, M., D'Alessio, 
P., Muzerolle, J., Calvet, N., \& Hartmann, L.\ 2003, \apj, 586, 1148 

\bibitem[{{Padgett} {et~al.}(2006){Padgett}, {Fukagawa}, {Rebull},
  {Noriega-Crespo}, {Carey}, {Stapelfeldt}, {Hillenbrand}, {Huard}, {Terebey},
  {Hines}, {Brooke}, {McCabe}, {Guedel}, {Knapp}, {Audard}, {Menard}, {Monin},
  {Dougados}, {Evans}, {Allen}, {Strom}, \& {Harvey}}]{padgett2006}
{Padgett}, D., et al. 2006, in Bulletin of the American Astronomical Society,
  Vol.~38, Bulletin of the American Astronomical Society, 947--+

\bibitem[{{Padgett} {et~al.}(1999){Padgett}, {Brandner}, {Stapelfeldt},
  {Strom}, {Terebey}, \& {Koerner}}]{padgett1999}
{Padgett}, D.~L., {Brandner}, W., {Stapelfeldt}, K.~R., {Strom}, S.~E.,
  {Terebey}, S., \& {Koerner}, D. 1999, \aj, 117, 1490

\bibitem[Pogge et al.(1998)]{tifkam} Pogge, R.~W., et al.\ 
1998, \procspie, 3354, 414 

\bibitem[{{Pollack} {et~al.}(1994){Pollack}, {Hollenbach}, {Beckwith},
  {Simonelli}, {Roush}, \& {Fong}}]{pollack1994}
{Pollack}, J.~B., {Hollenbach}, D., {Beckwith}, S., {Simonelli}, D.~P.,
  {Roush}, T., \& {Fong}, W. 1994, \apj, 421, 615

\bibitem[{{Ragan} {et~al.}(2006){Ragan}, {Bergin}, {Plume}, {Gibson}, {Wilner},
  {O'Brien}, \& {Hails}}]{ragan2006}
{Ragan}, S.~E., {Bergin}, E.~A., {Plume}, R., {Gibson}, D.~L., {Wilner}, D.~J.,
  {O'Brien}, S., \& {Hails}, E. 2006, \apjs, 166, 567

\bibitem[{{Rieke} {et~al.}(2004){Rieke}, {Young}, {Engelbracht}, {Kelly},
  {Low}, {Haller}, {Beeman}, {Gordon}, {Stansberry}, {Misselt}, {Cadien},
  {Morrison}, {Rivlis}, {Latter}, {Noriega-Crespo}, {Padgett}, {Stapelfeldt},
  {Hines}, {Egami}, {Muzerolle}, {Alonso-Herrero}, {Blaylock}, {Dole}, {Hinz},
  {Le Floc'h}, {Papovich}, {P{\'e}rez-Gonz{\'a}lez}, {Smith}, {Su}, {Bennett},
  {Frayer}, {Henderson}, {Lu}, {Masci}, {Pesenson}, {Rebull}, {Rho}, {Keene},
  {Stolovy}, {Wachter}, {Wheaton}, {Werner}, \& {Richards}}]{rieke2004}
{Rieke}, G.~H., et al. 2004, \apjs, 154, 25

\bibitem[{{Robitaille} {et~al.}(2007){Robitaille}, {Whitney}, {Indebetouw}, \&
  {Wood}}]{rob2007}
{Robitaille}, T.~P., {Whitney}, B.~A., {Indebetouw}, R., \& {Wood}, K. 2007,
  \apjs, 169, 328

\bibitem[{{Robitaille} {et~al.}(2006){Robitaille}, {Whitney}, {Indebetouw},
  {Wood}, \& {Denzmore}}]{rob2006}
{Robitaille}, T.~P., {Whitney}, B.~A., {Indebetouw}, R., {Wood}, K., \&
  {Denzmore}, P. 2006, \apjs, 167, 256

\bibitem[{{Rom{\'a}n-Z{\'u}{\~n}iga} {et~al.}(2007){Rom{\'a}n-Z{\'u}{\~n}iga},
  {Lada}, {Muench}, \& {Alves}}]{zuninga2007}
{Rom{\'a}n-Z{\'u}{\~n}iga}, C., {Lada}, C., {Muench}, A., \& {Alves}, J. 2007,
  ArXiv e-prints, 704

\bibitem[{{Seale} \& {Looney}(2008)}]{seale2007}
{Seale}, J.~P., \& {Looney}, L.~W. 2008, ApJ submitted

\bibitem[{{Shirley} {et~al.}(2000){Shirley}, {Evans}, {Rawlings}, \&
  {Gregersen}}]{shirley2000}
{Shirley}, Y.~L., {Evans}, II, N.~J., {Rawlings}, J.~M.~C., \& {Gregersen},
  E.~M. 2000, \apjs, 131, 249

\bibitem[{{Shu} {et~al.}(1987){Shu}, {Adams}, \& {Lizano}}]{shu1987}
{Shu}, F.~H., {Adams}, F.~C., \& {Lizano}, S. 1987, \araa, 25, 23

\bibitem[{{Simon} {et~al.}(2006){Simon}, {Jackson}, {Rathborne}, \&
  {Chambers}}]{simon2006}
{Simon}, R., {Jackson}, J.~M., {Rathborne}, J.~M., \& {Chambers}, E.~T. 2006,
  \apj, 639, 227

\bibitem[{{Skrutskie} {et~al.}(2006){Skrutskie}, {Cutri}, {Stiening},
  {Weinberg}, {Schneider}, {Carpenter}, {Beichman}, {Capps}, {Chester},
  {Elias}, {Huchra}, {Liebert}, {Lonsdale}, {Monet}, {Price}, {Seitzer},
  {Jarrett}, {Kirkpatrick}, {Gizis}, {Howard}, {Evans}, {Fowler}, {Fullmer},
  {Hurt}, {Light}, {Kopan}, {Marsh}, {McCallon}, {Tam}, {Van Dyk}, \&
  {Wheelock}}]{2mass}
{Skrutskie}, M.~F., et al. 2006, \aj, 131, 1163

\bibitem[{{Stark} {et~al.}(2006){Stark}, {Whitney}, {Stassun}, \&
  {Wood}}]{stark2006}
{Stark}, D.~P., {Whitney}, B.~A., {Stassun}, K., \& {Wood}, K. 2006, \apj, 649,
  900

\bibitem[Swift \& Welch(2008)]{swift2008} Swift, J.~J., \& Welch, W.~J.\ 2008, \apjs, 174, 202 

\bibitem[{{Tannirkulam} {et~al.}(2007){Tannirkulam}, {Harries}, \&
  {Monnier}}]{tann2007}
{Tannirkulam}, A., {Harries}, T.~J., \& {Monnier}, J.~D. 2007, \apj, 661, 374

\bibitem[{{Terebey} {et~al.}(1984){Terebey}, {Shu}, \& {Cassen}}]{tsc1984}
{Terebey}, S., {Shu}, F.~H., \& {Cassen}, P. 1984, \apj, 286, 529

\bibitem[{{Tobin} {et~al.}(2007){Tobin}, {Looney}, {Mundy}, {Kwon}, \&
  {Hamidouche}}]{tobin2007}
{Tobin}, J.~J., {Looney}, L.~W., {Mundy}, L.~G., {Kwon}, W., \& {Hamidouche},
  M. 2007, \apj, 659, 1404

\bibitem[{{Tuthill} {et~al.}(2001){Tuthill}, {Monnier}, \&
  {Danchi}}]{monnier2001}
{Tuthill}, P.~G., {Monnier}, J.~D., \& {Danchi}, W.~C. 2001, \nat, 409, 1012

\bibitem[{{Watson} \& {Stapelfeldt}(2007)}]{watson2007}
{Watson}, A.~M., \& {Stapelfeldt}, K.~R. 2007, \aj, 133, 845

\bibitem[{{Whitney} \& {Hartmann}(1993)}]{whitney1993}
{Whitney}, B.~A., \& {Hartmann}, L. 1993, \apj, 402, 605

\bibitem[{{Whitney} {et~al.}(2003{\natexlab{a}}){Whitney}, {Wood}, {Bjorkman},
  \& {Wolff}}]{whitney2003a}
{Whitney}, B.~A., {Wood}, K., {Bjorkman}, J.~E., \& {Wolff}, M.~J.
  2003{\natexlab{b}}, \apj, 591, 1049


\bibitem[{{Whitney} {et~al.}(2003{\natexlab{b}}){Whitney}, {Wood}, {Bjorkman},
  \& {Cohen}}]{whitney2003b}
{Whitney}, B.~A., {Wood}, K., {Bjorkman}, J.~E., \& {Cohen}, M.
  2003{\natexlab{a}}, \apj, 598, 1079



\bibitem[{{Whittet} {et~al.}(2001){Whittet}, {Gerakines}, {Hough}, \&
  {Shenoy}}]{whittet2001}
{Whittet}, D.~C.~B., {Gerakines}, P.~A., {Hough}, J.~H., \& {Shenoy}, S.~S.
  2001, \apj, 547, 872

\bibitem[{{Wood} {et~al.}(2001){Wood}, {Smith}, {Whitney}, {Stassun}, {Kenyon},
  {Wolff}, \& {Bjorkman}}]{wood2001}
{Wood} et al. 2001, \apj, 561, 299

\bibitem[{{Ybarra} {et~al.}(2006){Ybarra}, {Barsony}, {Haisch}, {Jarrett},
  {Sahai}, \& {Weinberger}}]{ybarra2006}
{Ybarra} et al. 2006, \apjl, 647, L159

\bibitem[{{Zhou} {et~al.}(1996){Zhou}, {Evans}, \& {Wang}}]{zhou1996}
{Zhou}, S., {Evans}, II, N.~J., \& {Wang}, Y. 1996, \apj, 466, 296




\end{thebibliography}

\clearpage

\clearpage

\clearpage

\clearpage

\clearpage

\clearpage

\clearpage

\clearpage

\clearpage

\clearpage

\clearpage

\clearpage

\clearpage

\begin{figure}
\figurenum{13b}
\begin{center}
\includegraphics[scale=.75]{f13b.eps}
\end{center}
\caption{}

\end{figure}

\clearpage

\begin{figure}
\figurenum{14}
\begin{center}
\includegraphics[scale=0.8]{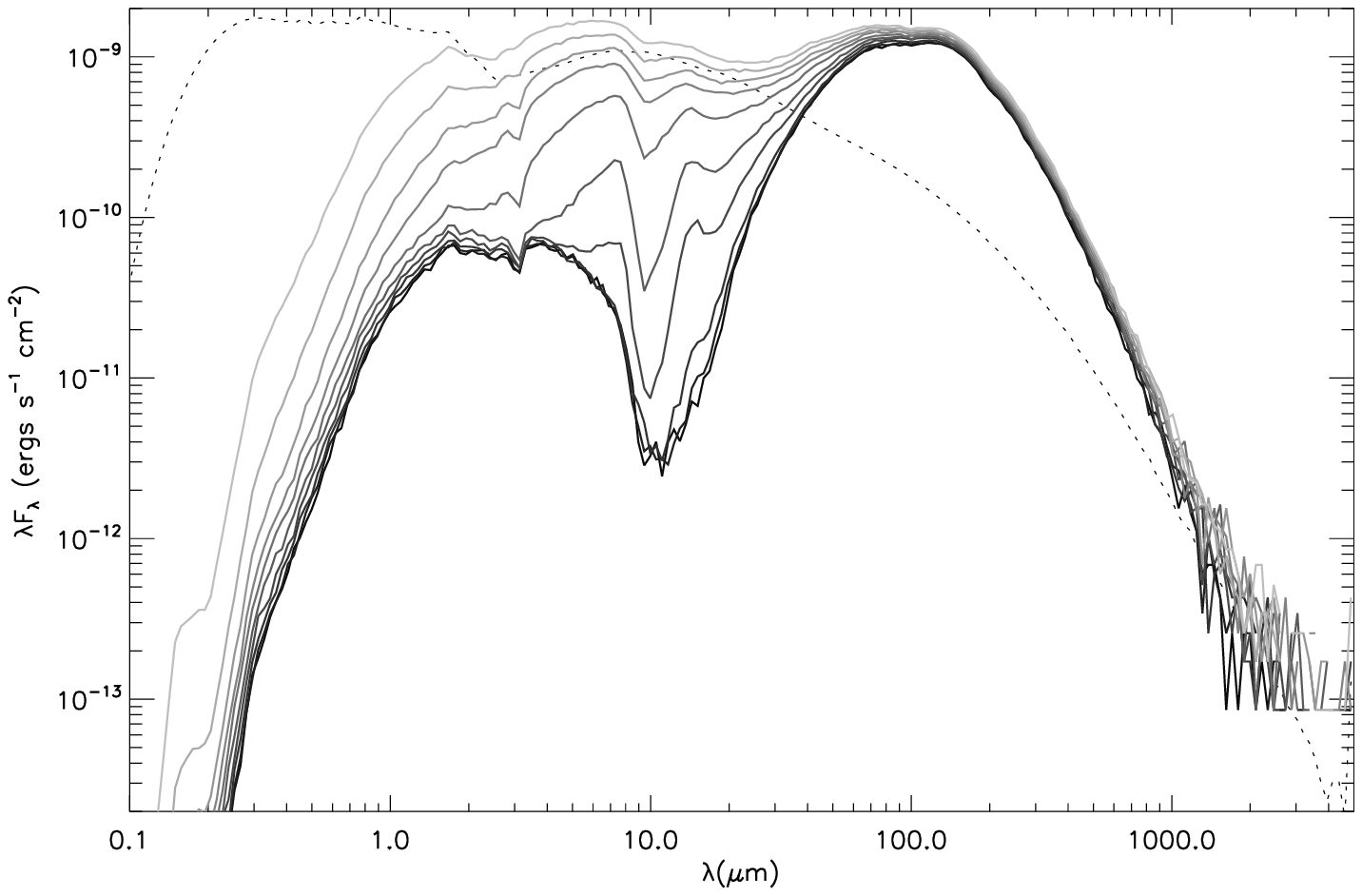}
\includegraphics[scale=0.8]{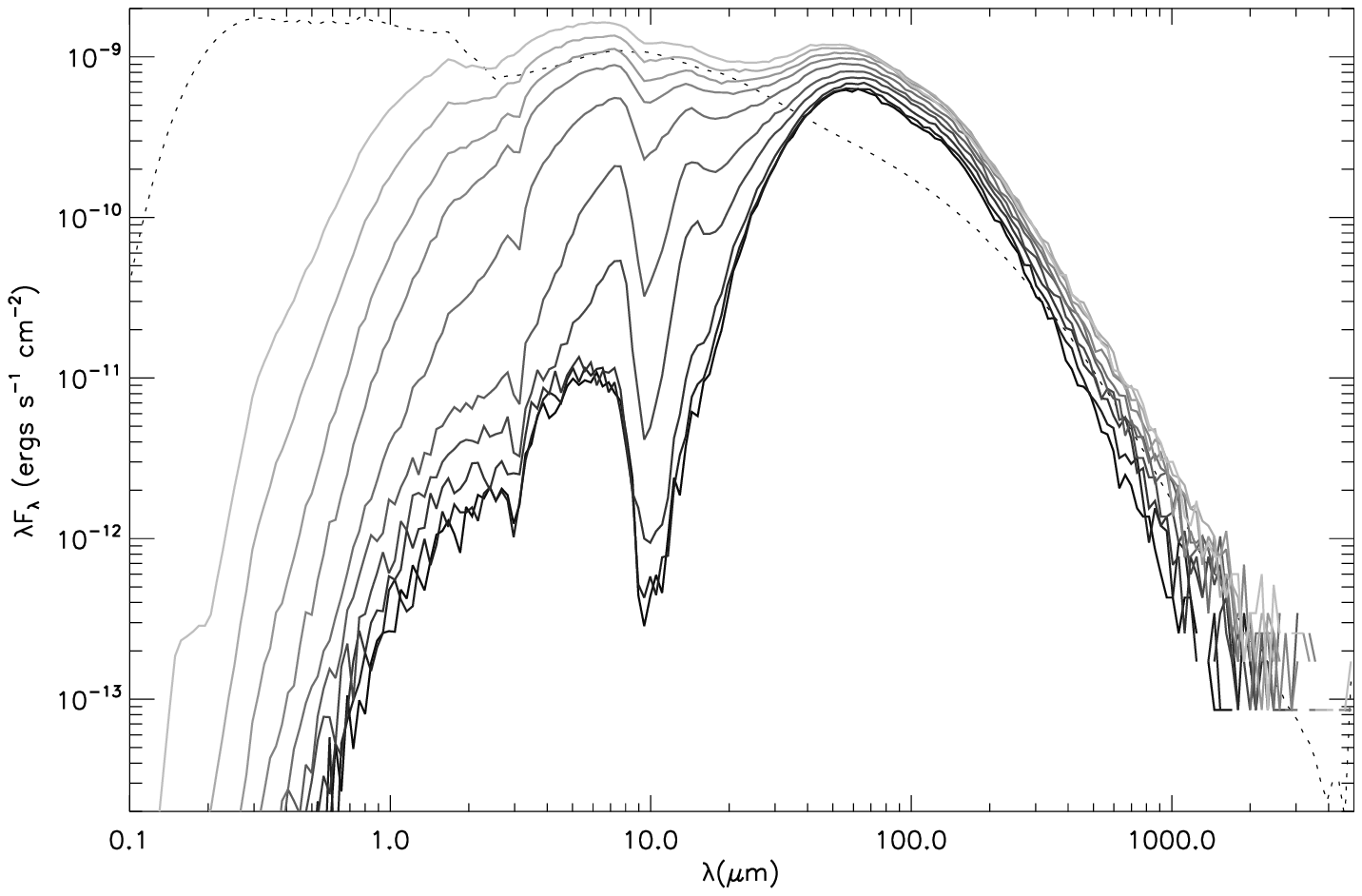}
\end{center}
\caption{SEDs of the best fitting model from L1527 at multiple
inclinations. The lines plotted from lightest to darkest correspond to
inclinations of 18, 32, 31, 49, 57, 63, 70, 76, 81, and 87$^{\circ}$ measured in
apertures of 5000 AU (left panel) and 1000 AU (right panel). The dotted line is the central star and disk 
as viewed face-on. }
\label{sedinc}
\end{figure}

\clearpage

\begin{figure}
\figurenum{15}
\begin{center}
\includegraphics[angle=-90, scale=0.8]{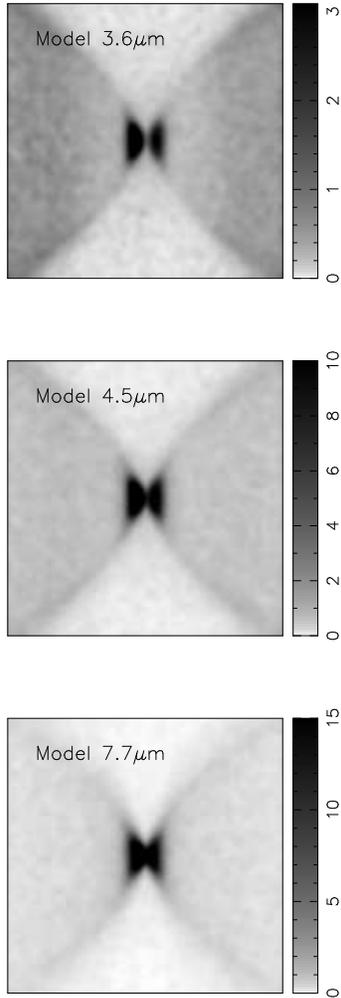}
\end{center}
\caption{Simulated images convolved with 0$\farcs$5 seeing. These
images are comparable to the abilities of current ground-based infrared telescopes. The 3.6$\mu$m image
is comparable to L$^{\prime}$-band, 4.5$\mu$m is comparable to M$^{\prime}$-band, and the 7.7$\mu$m
is comparable to a narrow-band filter.}
\label{simobs}
\end{figure}

\clearpage

\begin{deluxetable}{lllllll}

\tabletypesize{\small}
\tablecaption{Photometry\label{photo}}
\tablehead{
  \colhead{Wavelength} &  \colhead{F$_{\lambda}$} & \colhead{Aperture} & \colhead{Aperture}\tablenotemark{a} & \colhead{Instrument}  & \colhead{Observation} & \colhead{References}\\
  \colhead{($\mu$m)} &  \colhead{(mJy)} & \colhead{(arcsec)} & \colhead{Correction} & & \colhead{Date}&\\

   }
\startdata
1.66 & 7.0  $\pm$ 3.5 & 71.43 & -  & TIFKAM &  12-27-2007 & 1\\
1.66 & 8.35  $\pm$ 4.17 & 71.43 & -  & 2MASS &  10-14-1998 & 1\\
2.16 & 0.594 $\pm$ 0.162  & 7.143 & -  & TIFKAM & 12-27-2007 & 1\\
2.16 & 35.2 $\pm$ 16.2  & 71.43 & -  & TIFKAM & 12-27-2007 & 1\\
2.16 & 36.94 $\pm$ 24.3  & 71.43 & -  & 2MASS & 10-14-1998 & 1\\

3.6  & 6.936 $\pm$ 0.69 & 7.143 & 1.014  & IRAC  & 03-07-2004 & 1\\
" "  & 141.8 $\pm$ 16.2 & 71.43 & 0.9164  & IRAC  & 03-07-2004 & 1\\
3.6  & 6.132 $\pm$ 0.61 & 7.143 & 1.014  & IRAC  & 02-23-2005 & 1\\
" "  & 128.3 $\pm$ 22.6 & 71.43 & 0.9164  & IRAC  & 02-23-2005 & 1\\

4.5  & 22.75 $\pm$ 2.28 & 7.143 & 1.061  & IRAC  & 03-07-2004 & 1\\
" "  & 225.1 $\pm$ 16.3 & 71.43 & 0.9463  & IRAC  & 03-07-2004 & 1 \\
4.5  & 18.43 $\pm$ 1.84 & 7.143 & 1.061  & IRAC  & 02-23-2005 & 1\\
" "  & 190.0 $\pm$ 13.1 & 71.43 & 0.9463  & IRAC  & 02-23-2005 & 1 \\

5.8  & 29.93 $\pm$ 2.99 & 7.143 & 0.9918  & IRAC  & 03-07-2004 & 1\\
" "  & 149.5 $\pm$ 45.0 & 71.43 & 0.7926  & IRAC  & 03-07-2004 & 1\\
5.8  & 24.08 $\pm$ 2.41 & 7.143 & 0.9918  & IRAC  & 02-23-2005 & 1\\
" "  & 163.4 $\pm$ 45.5 & 71.43 & 0.7926  & IRAC  & 02-23-2005 & 1\\
6.7  & 22.36 $\pm$ 5.6  & 7.2   &  -      & ISOCAM & 10-02-1997 & 1\\
" "  & 62.68 $\pm$ 15.7 & 36    &  -      & ISOCAM & 10-02-1997 & 1\\

8.0  & 18.83 $\pm$ 3.8  & 7.143 & 0.9423  & IRAC  & 03-07-2004 & 1\\
" "  & 54.54 $\pm$ 25.0 & 71.43 & 0.7629  & IRAC  & 03-07-2004 & 1\\
8.0  & 14.87 $\pm$ 3.0  & 7.143 & 0.9423  & IRAC  & 02-23-2005 & 1\\
" "  & 37.43 $\pm$ 18.0  & 71.43 & 0.7629  & IRAC  & 02-23-2005 & 1\\
9.6  & 1.37  $\pm$ 0.34 & 7.2   &  -      & ISOCAM & 10-02-1997 & 1\\
11.3 & 3.081 $\pm$ 0.77 & 7.2   &  -      & ISOCAM & 10-02-1997 & 1\\
14.3 & 15.3  $\pm$ 3.8  & 7.2   &  -      & ISOCAM & 10-02-1997 & 1\\
" "  & 54.76 $\pm$ 13.7 & 36    &  -      & ISOCAM & 10-02-1997 & 1\\
24   & 660.6 $\pm$ 66 & 13    & 1.167  & MIPS  & 03-05-2005 & 1\\
25   & 743.6 $\pm$ 70 & 45 x 300  & -  & IRAS & - & 2\\
60   & 17770 $\pm$ 1600 & 90 x 300 & -  & IRAS & - & 2\\
70   & 24170 $\pm$ 4834 &  75   & 1.0  & MIPS  & 03-05-2005 & 1\\
100  & 73260 $\pm$ 11700 & 180 x 300 & -  & IRAS & - & 2\\
100  & 89000 $\pm$ 36000 & 60 & -  & Yerkes (KAO)\tablenotemark{b} & 03-1987 & 3\\
160  & 94000 $\pm$ 38000 &  60    & -  & Yerkes (KAO) & 03-1987 & 3\\
350  & 22000 $\pm$ 9000  &  60    & -  & Yerkes (IRTF) & 10-1987 & 3\\
350  & 66000 $\pm$ 20000 & 45    & -  & SCUBA  & 01-1998 & 4\\
450  & 14000 $\pm$ 5600 & 60    & -  & JCMT/UKT  & 01-1989 & 4\\
450  & 44800 $\pm$ 9000 & 45    & -  & SCUBA  & 01-1998 & 4\\
450  & 18200 $\pm$ 3200 & 40   & -   & SCUBA  & 01-1998 & 5\\
450  & 55500 $\pm$ 20900 & 120   & -   & SCUBA  & 01-1998 & 5\\
750  & 8400 $\pm$ 1100 & 45    & -  & SCUBA  & 01-1998 & 4\\
800  & 1400 $\pm$ 560 & 60     & -  & JCMT/UKT & 01-1989 & 3\\
850  & 5900 $\pm$ 480 & 45    & -  & SCUBA  & 01-1998 & 4\\
850  & 3190 $\pm$ 190 & 40    & -  & SCUBA  & 01-1998 & 5\\
850  & 9410 $\pm$ 460 & 120    & -  & SCUBA  & 01-1998 & 5\\
1300 & 720 $\pm$ 110 & 40    & -  & SCUBA  & 01-1998 & 5\\
1300 & 1500 & 30    &   -  & IRAM    & - & 6\\
2700 & 47$\pm$ 5.6 & 60 x 60 & -  & NMA\tablenotemark{c} & 01-1995 & 7 
\enddata
\tablenotetext{a}{Corrections for IRAC data were derived from the prescription for extended source calibration on the \textit{Spitzer Science Center} website. Corrections for MIPS data are taken from the MIPS data handbook.}
\tablenotetext{b}{Kuiper Airborne Observatory}
\tablenotetext{c}{Nobeyama Millimeter Array}
\tablerefs{(1) This work; (2) \citet{iras}; (3) \citet{ladd1991}; (4) \citet{chandler2000}; (5) \citet{shirley2000}; (6) \citet{motte2001}; (7) \citet{ohashi1997}}
\end{deluxetable}

\begin{deluxetable}{llcccccc}
\tabletypesize{\scriptsize}
\rotate
\tablewidth{0pt}
\tablecaption{Model parameters\label{param}}
\tablehead{
  \colhead{Parameter} & \colhead{Description} & \colhead{This paper}  & \colhead{KCH93}  & \colhead{R07 Standard} & \colhead{R07 Full SED} & \colhead{R07 IRAC SED} & \colhead{Furlan 2007}\\
 & & & & Integrated & Resolved & Resolved &
 }
\startdata
R$_{*}$(\rsun) & Stellar radius & 2.09 & - & - & 8.21 & 21.57 & 2.0\\
T$_{*}$(K) & Stellar temperature & 4000 & - & 2932 - 3869 & 4260 & 4360 & 4000\\
L$_{*}$(L$_{\sun}$) & System luminosity & 2.75 & 1.35 & 0.64 - 3.84 & 20 & 155 & 1.8\\
M$_{*}$(M$_{\sun}$) & Stellar mass & 0.5 & - & 0.15 - 0.59 & 1.46 & 4.07 & -\\
M$_{disk}$(M$_{\sun}$) & Disk mass & 0.1 & - & 6.3 $\times 10^{-5}$ - 0.016 & 8.58 $\times 10^{-4}$ & 0.142 & - \\
h(100) (AU) & Disk scale height at 100AU & 10.52 &- & 3.18 - 9.67 & 8.09 & 3.891 & 10 \\
$\alpha$ & Disk radial density exponent & 2.125 & - & - & 2.155 & 2.042 &- \\
$\beta$ & Disk scale height exponent & 1.125 & - & - & 1.155 & 1.042 & -\\
$\dot{M}_{disk}$(M$_{\sun}$ $yr^{-1}$) &  Disk accretion rate & 3.0 $\times 10^{-7}$ & -& 9.77 $\times 10^{-12}$ - 5.01 $\times 10^{-8}$ & 1.98 $\times 10^{-8}$ & 7.29 $\times 10^{-7}$ & - \\
R$_{trunc}$(R$_{*}$) & Magnetosphere co-rotation radius & 3.0 & - & - & 5.0 & 5.0 & -\\
F$_{spot}$ & Fractional area of accretion hotspot & 0.01 & - & 0.01 & 0.01 & 0.01 & -\\
R$_{disk,min}$(R$_{*}$) & Disk inner radius & 14.25 & - & R$_{dd}$ & 7.84 & 8.33 & 1.0\\
R$_{disk,max}$(AU) & Disk outer radius & 75 & - & 33.8 - 1303.3 & 58.398 & 33.703 & 200 \\
R$_{c}$(AU) & Centrifugal radius & 75 & 300 & - & 58.398 & 33.703 & 200\\
R$_{env,min}$(R$_{*}$) & Envelope inner radius & 42.75 & - & - & 7.84 & 8.33 & -\\
R$_{env,max}$(AU) & Envelope outer radius & 15000 & 3000 & - & 16200 & 9120 & 10000\\
$\dot{M}_{env}$(M$_{\sun}$ $yr^{-1}$) & Envelope mass infall rate & 1.00 $\times 10^{-5}$ & 4.22 $\times 10^{-5}$\tablenotemark{a} & 0.911 - 5.05 $\times 10^{-5}$ & 1.37 $\times 10^{-4}$ & 2.63 $\times 10^{-4}$ & 5.34 $\times 10^{-6}$\tablenotemark{a}\\
$\rho_1$(g cm$^{-3}$) & Envelope Density at 1 AU & 3.75 $\times 10^{-14}$ & 3.16 $\times 10^{-13}$ & 0.341 - 1.89 $\times 10^{-13}$\tablenotemark{a} & 8.77 $\times 10^{-13}$ & 2.81 $\times 10^{-12}$ & 1.00 $\times 10^{-14}$\\  
b$_{in}$ & Inner cavity shape exponent & 1.5 & - & - & - & - & \\
b$_{out}$ & Outer cavity shape exponent & 1.9 & none & 1.5 & 1.5 & 1.5 & streamline\\
z$_{in}$(AU) & Inner cavity offset & 0 & - & - & - & - & -\\
z$_{out}$(AU) & Outer cavity offset height & 100 & - & - & - & - & - \\
$\theta_{open,in}$($^{\circ}$) & Inner cavity opening angle & 15 & - & - & - & - & -\\
$\theta_{open,out}$($^{\circ}$) & Outer cavity opening angle & 20 & - & - & 43 & 16 & 27\\
$\theta_{inc}$($^{\circ}$) & Inclination angle & 85 & 60-90 & 41-81 & 81 & 75 & 89\\
$\rho_{c}$(g cm$^{-3}$) & Cavity density & 0 & - & - & 1.40 $\times 10^{-20}$ & 1.42 $\times 10^{-20}$ & -\\
$\rho_{amb}$(g cm$^{-3}$) & Ambient density & 0 & - & - & 3.838 $\times 10^{-22}$ & 1.42 $\times 10^{-20}$ & -

\enddata
\tablenotetext{a}{Assumes a 0.5\msun central stellar mass.}
\end{deluxetable}

\clearpage

\clearpage

\end{document}